\documentclass[12pt,preprint]{aastex}
\usepackage{graphics,graphicx,epsfig,lscape}

\usepackage{amssymb}
\usepackage{epstopdf}
\usepackage{natbibspacing}
\usepackage{natbib}
\newcommand{\msun}{M$_{\odot}$ }

\newcommand{\ie}{\textit{i.e.}~}
\newcommand{\GALEX}{\textit{GALEX}~}

\newcommand{\HST}{\textit{HST}~}

\begin{document}

\title{Studying Large and Small Scale Environments of Ultraviolet Luminous Galaxies}
\author{
Antara R. Basu-Zych\altaffilmark{1}, 
David Schiminovich\altaffilmark{1}, 
Sebastien Heinis\altaffilmark{7}, 
Roderik Overzier\altaffilmark{11}
Tim Heckman\altaffilmark{7}, 
Michel Zamojski\altaffilmark{2}, 
Olivier Ilbert\altaffilmark{12},
Anton M. Koekemoer\altaffilmark{13},
Tom A. Barlow\altaffilmark{2},
Luciana Bianchi\altaffilmark{7},
Tim Conrow\altaffilmark{2},
Jose Donas\altaffilmark{5},
Karl G. Forster\altaffilmark{2},
Peter G. Friedman\altaffilmark{2},
Young-Wook Lee\altaffilmark{8},
Barry F. Madore\altaffilmark{4},
D. Christopher Martin\altaffilmark{2},
Bruno Milliard\altaffilmark{5},
Patrick Morrissey\altaffilmark{2},
Susan G. Neff\altaffilmark{9},
R. Michael Rich\altaffilmark{10},
Samir Salim\altaffilmark{3},
Mark Seibert\altaffilmark{4},
Todd A. Small\altaffilmark{2},
Alex S. Szalay\altaffilmark{6},
Ted K. Wyder\altaffilmark{2},
Sukyoung Yi\altaffilmark{7}}

\altaffiltext{1}{Department of Astronomy, Columbia University, 550 West 120th Street, New York, NY 10027; antara@astro.columbia.edu}
\altaffiltext{2}{California Institute of Technology, MC 405-47, 1200 East California Boulevard, Pasadena, CA 91125}
\altaffiltext{3}{NOAO, Tuscon, Arizona}
\altaffiltext{4}{Observatories of the Carnegie Institution of Washington, 813 Santa Barbara St., Pasadena, CA 91101}
\altaffiltext{5}{Laboratoire d'Astrophysique de Marseille, BP8, Traverse du Siphon, F-13376 Marseille, France}
\altaffiltext{6}{Department of Physics and Astronomy, The Johns Hopkins University, Homewood Campus, Baltimore, MD 21218}
\altaffiltext{7}{Center for Astrophysical Sciences, The Johns Hopkins` University, 3400 N. Charles St., Baltimore, MD 21218}
\altaffiltext{8}{Center for Space Astrophysics, Yonsei University, Seoul 120-749, Korea}
\altaffiltext{9}{Laboratory for Astronomy and Solar Physics, NASA Goddard Space Flight Center, Greenbelt, MD 20771}
\altaffiltext{10}{Department of Physics and Astronomy, University of California, Los Angeles, CA 90095}
\altaffiltext{11}{Max-Planck-Institut f\"{u}r Astrophysik, D-85748 Garching, Germany}
\altaffiltext{12} {Institute for Astronomy, 2680 Woodlawn Dr., University of Hawaii, Honolulu, Hawaii, 96822}
\altaffiltext{13}{Space Telescope Science Institute, 3700 San Martin Drive, Baltimore, MD 21218}

\begin{abstract}
Studying the environments of 0.4$<$z$<$1.2 UV-selected galaxies, as examples of extreme star-forming galaxies (with star formation rates in the range of 3$-$30\msun yr$^{-1}$), we explore the relationship between high rates of star-formation, host halo mass and pair fractions. We study the large-scale and small-scale environments of local Ultraviolet Luminous Galaxies (UVLGs) by measuring angular correlation functions. We cross-correlate these systems with other galaxy samples: a volume-limited sample (ALL), a Blue Luminous Galaxy sample (BLG) and a Luminous Red Galaxy sample (LRG). We determine the UVLG comoving correlation length to be r$_0=4.8 ^{+11.6}_{-2.4}$ h$^{-1}$Mpc at $\langle$z$\rangle =$1.0, which is unable to constrain the halo mass for this sample. However, we find that UVLGs form close (separation $<$ 30 kpc) pairs with the ALL sample, but do not frequently form pairs with LRGs. A rare subset of UVLGs, those with the highest  FUV surface brightnesses, are believed to be local analogs of high redshift Lyman Break Galaxies (LBGs) and are called Lyman Break Analogs (LBAs). LBGs and LBAs share similar characteristics (i.e., color, size, surface brightness, specific star formation rates, metallicities, and dust content). Recent \HST images of z$\sim$0.2 LBAs show disturbed morphologies, signs of mergers and interactions. UVLGs may be influenced by interactions with other galaxies and we discuss this result in terms of other high star-forming, merging systems. 
\end{abstract}
\section{Introduction}
The observed properties of galaxies can be attributed to a wide variety of complex factors. In order to understand the environmental causes that might trigger intense star-formation in the local universe, such as that observed at high redshifts for Lyman Break Galaxies, we explore two regimes: large-scale, to study the connection between halo mass and high star formation rates (SFRs); and small scale, calculating merger rates. At large scales, a galaxy's SFR may be connected to how the galaxy accretes gas from its large-scale environment, which depends on the halo mass. Smaller scale environmental effects, from nearby galaxies, may induce star formation from interactions and eventual mergers.

The study of galaxy evolution has been connected to that of larger structure formation-- their growth related to overdensities arising from primordial density fluctuations. Studies of galaxy bias, the distribution of galaxies relative to underlying dark matter distribution, attempt to describe the fate and origin of present-day galaxies. The link between halo mass and star formation has been discussed and explored in several papers \citep{wang06, mandelbaum06, conroy08, khochfar}. A simple description involves gas flowing to centers of dark matter potentials, reaching sufficient density to radiate and cool, losing pressure support and settling into a disk, and then forming stars. The system is eventually accreted onto a larger structure. Such accretion by clusters can shock-heat the gas into high temperatures, preventing new stars from forming. Therefore, we expect that active star formation can occur for halo masses whose virial temperatures are less than 10$^{7}$K, where shock-heating inhibits new star formation. 

Redshift evolution of the link between halo mass and star formation, determined by large scale environmental effects, might explain observations such as the Butcher-Oemler \citep{Oemler74} effect. In this effect, clusters at high redshifts host more blue, star-forming galaxies than local clusters, which are predominantly inhabited by ``red and dead'' galaxies.  Similarly, using the two-point angular correlation function to measure clustering strength, \cite{Heinis} find that UV-selected, star-forming galaxies have migrated from cluster regions, with high correlation lengths (r$_0$), at high redshift to more isolated regions at low redshifts. From the view of hierarchical structure formation, higher redshift clusters have lower halo masses compared to low-z clusters, thus high-z clusters have virial temperatures low enough to allow star formation. Adding to this framework, \cite{Keres} discuss the importance of hot vs. cold accretion -- in the simplified picture above, gas does not get heated to the virial temperature necessarily. Massive galaxies draw gas via quasi-spherical, ``hot mode'' accretion, in which the gas does shock-heat to the virial temperature. However, less massive galaxies acquire gas from flow along filaments in ``cold mode'' accretion where the gas radiates away its gravitational energy and never reaches the virial temperature (see \cite{Keres, Croton} and references therein). These simulations show that the large-scale galaxy environment influences gas accretion processes. Therefore, a galaxy's halo mass contributes significantly to how it acquires gas, the fuel that affects its star formation efficiency. 

Similar shock physics may affect the relation between small scale merger effects and star formation. \cite{KWG} predict that the merger of a halo with a larger structure, such as a cluster, suppresses star formation as gas is shock-heated to the virial temperature of the more massive halo and then cools onto the central object of the most massive halo. Thus, these results suggest that the mass of the larger companion is a consideration for whether mergers might enhance star formation. According to the work of \cite{KWG}, mergers with clusters should not result in enhanced star formation. 

Nevertheless, other studies show that at these small scales galaxy-galaxy interactions and close neighbors can stimulate star formation activity. In hierarchical structure formation, the merger process is an important ingredient for galaxy evolution. Over thirty years ago, \cite{Toomre2} suggested that tidal forces during the merger process can funnel gas into the central regions of galaxies, triggering strong star formation activity. Further theoretical studies have investigated the connection between mergers and high star formation rates (see \cite{mihos94}, \cite{cox}, and \cite{diMatteo}). 

\cite{Li07} have shown that nearly half of the highest star-forming galaxies have close companions (galaxies within a projected radius of 100 kpc), by correlating 10$^5$ galaxies from the Sloan Digital Sky Survey (SDSS). According to this research, tidal interactions are the primary cause for enhanced star formation for these most active star-forming galaxies. Several studies using various star formation indicators from H$\alpha$, CO, far-infrared luminosities for interacting galaxies have shown that enhanced star formation results from galaxy interactions (e.g., \cite{keel}, \cite{Struck}, and references therein). 

Using the Galaxy Evolution Explorer (\GALEX), \cite{Heckman05} have uncovered a population of local, intensely star-forming galaxies. These galaxies are selected based on their FUV luminosities (L$_{\rm FUV}$ $>$ 2$\times$10$^{10} L_{\odot}$). The most compact of these UV-luminous galaxies (UVLGs), with I$_{\rm FUV} >10^{9}$ L$_{\odot}$ kpc$^{-2}$, share many similar properties with z$\sim$3 LBGs: specific SFRs, metallicities, morphologies, kinematics, and attenuations \citep{choopes, rod, me2}. We refer to these systems as Lyman Break Analogs (LBAs), to discriminate this special subset of UVLGs from lower surface brightness UVLGs. Given that LBGs are found at higher redshifts (z$=$2-3) where major and minor mergers are generally more prevalent than for z$=$0.1 LBAs, LBAs and LBGs may be expected to differ in their recent merger histories, environments, and location within large scale structure. 

In this paper, we study the environments, as quantified by correlation functions, of low to intermediate redshift (0.4$<$z$<$1.2) UVLGs in the COSMOS field. We extend our analysis to LBAs to compare with z$\sim$3 LBGs, although this analysis is limited because of the rarity of z$<$1 LBAs. We measure two-point auto-correlation functions (ACFs) and cross-correlation functions (CCFs) of UVLGs with other galaxy samples to probe the mass of the dark halos within which UVLGs reside. We connect these observations with those of LBGs at high redshift (z$\sim$3), which are known to strongly cluster \citep{adel05}. Recent Hubble Space Telescope (\HST) images of z$\sim$0.2 LBAs show disturbed morphologies, signs of mergers and interactions \citep{rod}. We relate small-scale correlations to pair fractions in order to estimate merger rates. Studying the locations of current UVLGs, as examples of extreme star-forming galaxies, we explore the relationship between high rates of star-formation, host halo mass, and merger rate as a function of redshift.

In the following section, we discuss our sample selection. In the three subsequent sections, we outline the theory (\S\ref{sec:theory}) and techniques for measuring the angular correlation functions(\S\ref{sec:meas}), and derivations of physical quantities (\S\ref{sec:quant}) used to analyze our data. We discuss our results for the large and small scale environments of UVLGs and LBAs in \S\ref{sec:results}, and discuss these in the context of other environmental studies in \S\ref{sec:disc}. A summary of our analysis can be found in the final section: \S\ref{sec:conc}. Throughout this paper we adopt the cosmology given by (H$_0$, $\Omega_{\rm M}$, $\Omega_{\Lambda}$) = (70 km s$^{-1}$ Mpc$^{-1}$, 0.3, 0.7). Unless explicitly stated, our number densities are in physical, proper units. Correlation lengths are quoted in comoving units of h$^{-1}$Mpc.

\section{Data and Sample Selection} \label{sec:data}
With its large contiguous field and deep observations, the COSMOS survey provides excellent data for environmental studies. The COSMOS field covers 2 deg$^{2}$, centered on $\alpha_{\rm J2000} = 10^{h} 00^{m} 28.6^{s}$, $\delta_{\rm J2000} = +02^{\circ} 12\arcmin 21.0 \arcsec$. This region of sky has rich multiwavelength data, including: optical observations with the superb Subaru ground-based telescope, atop Mauna Kea, Hawaii, ultraviolet coverage from \GALEX, radio observations from VLA, and Xray data from XMM. The inner 1.7 deg$^{2}$ has been imaged by \HST, using the Advanced Camera Survey (ACS) in the F814W filter (see \cite{Anton} for details). Deep photometric data from Subaru provides photometric redshifts with $\Delta \rm{z}/\rm{(1+z)} \le 0.02$, depending on I-band magnitude and redshift \citep{Ilbert}. 

From the I-band selected (I$<$25.5) COSMOS photometric redshift catalog (for details about this catalog see \cite{Capak07}), we compare UVLG and LBA samples, which draws from the combined COSMOS+\GALEX data (I$<$25.5, and NUV$<$23, See \cite{Zamojski} for more details), with three separate galaxy samples (selected from the COSMOS sample, with I$<$25.5) for 0.4$<$z$<$1.2: (1) a volume limited sample in M$_{\rm G}$-``ALL'', (2) Blue Luminous Galaxies- ``BLG'', and (3) Luminous Red Galaxies- ``LRG''. We briefly note here -- while the ``ALL'' sample is volume-limited in M$_{\rm G}$, it is not volume-limited for all colors. This is discussed in more detail further below. Figure 1 shows the various cuts applied to the data to create these separate samples:  ALL is comprised of galaxies with -24$<{\rm M}_{\rm g}<$-18; BLG galaxies have ${\rm M_g}<$-20.0 and ${\rm M_u}-\rm{M_r}<$ 1.8 color; and LRG set is selected by a luminosity cut, ${\rm M_r}<$-20.5 and color cut, ${\rm M_u}-{\rm M_r}$ $>$ 2.0. Both colors and luminosities are given in rest-frame, k-corrected AB magnitudes.  K-corrections were done using the {\em kcorrect\_4.1.4} IDL routine \citep{kcorrect}. The spectral energy distribution (SED) fitting was done using the COSMOS optical bands as well as the \GALEX FUV and NUV bands, where available-- since the comparison samples are not required to be UV-detected. We modified this routine to use the closest band to the rest-frame filters for determining kcorrections-- for example, at z$=$0.5, the observed NUV magnitude at 2300\AA~best approximates the rest-frame FUV magnitude at 1500\AA (i.e, $2300\AA=1500\AA*(1+0.5)$). 

The UVLGs are selected using their k-corrected, rest-frame FUV properties from the COSMOS+\GALEX catalog: FUV $\le$-19.54, corresponding to the luminosity criterion of L$_{\rm FUV} > 10^{10.3} L_\odot$ \citep{Heckman05}. To estimate the FUV surface brightness, we use the value of r$_{50}$ output from SExtractor \citep{sex} for the ACS I-band data (see \cite{Zamojski} for further detail), which corresponds to the half-light radius. We use this ACS I-band half-light radius as a proxy for the UV size, acknowledging that the I-band at these redshifts corresponds to optical wavelengths. Our sample of LBAs were determined by the criteria: I$_{FUV} \ge10^9$ L$_\odot$ kpc$^{-2}$. While the I-band size may overestimate the UV size by a factor of 2 (\cite{rod}, thereby, underestimating the FUV surface brightness by a factor of 4), we find that $\sim 20$\% of UVLGs are LBAs, consistent with what is found at z$\sim$ 0.2 \citep{choopes}. 

We summarize the sample sizes for each redshift bin in Table 1, and show the redshift and spatial distributions of these samples in Figures 2 and 3, respectively. 

The ``L'' shape of the color-magnitude diagrams in Figure 1 indicates that the red and blue galaxies do not populate the magnitude space in the same way. The majority of red galaxies tend to be more luminous than blue galaxies; the faint end for the red galaxies starts decreasing around M$_{\rm g}=$-19. While the volume-limited cut shown on the upper left panel of Figure 1 selects a complete sample in M$_{\rm g}$, the color-magnitude diagram at M$_{\rm g} \sim$-19 shows a diminishing sample of red galaxies. The luminosity function (LF) for red galaxies shows that the faint end does have a decreasing slope (\cite{Zucca, Faber}, but our LF comparison of the red galaxies in the ALL sample with the theoretical LF does indicate sample incompleteness at higher redshifts. This incompleteness does not affect either the BLG or LRG samples which are selected to be sufficiently bright to avoid magnitude limits. However, the ALL magnitude cut of M$_{\rm g}=$-18 suffers from some incompleteness in the red population. This observation is worth noting for interpreting and comparing the ALL sample to other samples and studies, but does not affect the overall conclusions of this paper.

The predicted n(z), shown as a black dotted line in Figure \ref{fig:zdist}, is derived using the luminosity functions observed by \cite{Faber} for the ALL, BLG and LRG samples and by \cite{Arnouts04} for the UVLG sample (this is discussed further in \S\ref{sec:quant}). The redshift distributions in Figure 2 roughly follow the smooth predicted distribution marked by the black dotted line. The UVLG, as well as the LBA, redshift distributions may be too noisy, because of poor number statistics, to interpret their shape. The overdensities apparent in the spatial maps (Figure \ref{fig:skydist}) correspond to known physical structures observed in the COSMOS field (see \cite{NS07}, \cite{finoguenov}, \cite{Massey07} for more discussion of these structures). 

\subsection{Correlation Functions -- the Theory} \label{sec:theory}
In the standard framework for defining the two-point correlation function, galaxies are distributed in space such that the probability, $\delta \rm{P}$, of finding a galaxy in some volume, $\delta \rm{V}$ is, $\delta \rm{P} = {\it n}\delta \rm{V}$, where {\it n} is the mean number density of galaxies. Extrapolating to the situation of finding two galaxies at some comoving separation, r$_{12}$, one within volume, $\delta {\rm V}_1$ and another within volume, $\delta {\rm V}_2$, is then 
\begin{equation}
\delta \rm{P} = {\it n}^2\delta \rm{V}_1 \delta \rm{V}_2 [1+\xi (\rm{r}_{12})] \label{eqn:xi}
\end{equation}
where $\xi (\rm{r}_{12})$ is defined as the two-point correlation function. An approximation for the correlation function, $\xi(r)$, is a power-law with a single slope at all scales: 
\begin{equation}
\xi{\rm(r, z)} =(\rm{r/r_0(z)})^{-\gamma}  \label{eqn:xiform} % {\rm (1+z)^{-(3+\epsilon)}(r/r_0)^{-\gamma}}  \label{eqn:xiform}
\end{equation}
where z refers to redshifft, $\gamma$ and ${\rm r_0(z)}$ specify the slope and comoving clustering strength, respectively. Given a uniform, Poissonian distribution, $\xi({\rm r}) = 0$, while correlated data ($\xi({\rm r}) > 0$) means that there is an increased probability of finding a second galaxy within distance r of the first. 

Various studies (i.e., \cite{Zehavi04} for SDSS galaxies and \cite{Ouchi05b} for high-z LBGs) have found departures from a power-law with a single slope. Rather, they find that the correlation function is better approximated by the sum of two contributions: the ``one-halo term'', dominating at small scales for galaxy pairs residing within the same halo, and the ``two-halo term'', dominating at large scales for galaxy pairs within separate haloes. This so-called HOD modeling is beyond the scope of this paper. However, to account for the fact that the galaxy correlation function might not follow a power law with a single slope, we assume that $\xi (\rm{r})$ can be fit by two power laws with different slope for small scales and for large scales. 

For data based on photometric redshifts, the three dimensional correlation function can not be measured directly, but it can be inferred from the projected two-dimensional angular correlation function. By analogy of the steps outlined above, the three-dimensional, real space $\xi({\rm r})$ is collapsed into a two-dimensional function: w($\theta$), which measures the excess probability that two galaxies are found within some angular separation, $\theta$. 
\begin{equation}
\delta\rm{P} = {\it N}^2 [1+{\rm w(\theta)}]\delta\Omega_1\delta\Omega_2 \label{eqn:wtheta}
\end{equation}
where {\it N} is the surface density and $\delta\Omega_1$ and $\delta\Omega_2$ are the solid angle elements describing our field. Like $\xi(\rm r)$ in Eqn. \ref{eqn:xiform}, w($\theta$) is also expected to have a similar power law form:
\begin{equation}
\rm{w(\theta)} = {\rm a_{w,\theta}} \theta^{-\delta}. \label{eqn:wthetaform}
\end{equation}
We convert the three-dimensional correlation function into the angular one, using the formalism introduced by \cite{Limber} and \cite{Rubin}. Therefore, we relate the comoving correlation length, $\rm{r_0}$, to ${\rm a_{w,\theta}}$ and $\gamma$ to $\delta$ by the following:
 \begin{eqnarray}
 \gamma &=& \delta + 1\\
 \rm a_{w, \theta}&=&{\rm H}_\gamma \frac{\int_{\rm z}{\rm r_0(z)}^\gamma {\rm g(z)} (n_1 n_2) {\rm dz}}{\int_{\rm z} n_1({\rm z}) {\rm dz} \int_{\rm z} n_2({\rm z}) \rm dz} \label{eqn:r0}\\
\rm g(z)&=& \left ( \frac{\rm dz}{\rm dr}\right ) \rm r^{(1-\gamma)} \rm F(r)
 \end{eqnarray}
 where ${\rm H_\gamma} = \int _{-\infty}^{+\infty} {\rm dx (1+x^2)^{-\gamma/2}} = \frac{\Gamma(\onehalf)\Gamma(\frac{\gamma-1}{2})}{\Gamma(\frac{\gamma}{2})}$, ${\rm r}$ is the comoving distance, $n_1(z)$ and $n_2(z)$ are the number densities, and F(r) is the curvature term from the Robertson-Walker metric, 
 \begin{equation}
 {\rm ds^2= c^2 dt^2 - a^2[dr^2/F(r)^2 + r^2 d\theta^2 + r^2 sin^2\theta d\phi^2]}
 % \rm{r_0}&=& \left[ \frac{a_{w, \theta} }{{\rm H_\gamma H_0}\int_{\rm z }n_1 n_2 {\rm (1+z)^{p} r_c^{1-\gamma} [\Omega_\Lambda+\Omega_M(1+z)^3]^{1/2}  dz}}\right ]^{1/\gamma} \label{eqn:r0}\\ 
 %\rm{r_0}&=& \left[ \frac{a_{w, \theta} }{{\rm H_\gamma H_0}\int_{\rm z }n_1 n_2 {\rm (1+z)^{p} r_c^{1-\gamma} [\Omega_\Lambda+\Omega_M(1+z)^3]^{1/2}  dz}}\right ]^{1/\gamma} \label{eqn:r0}\\ 
 \end{equation}
Using the \cite{GP} form for the evolution of ${\rm r_0(z)}$:
\begin{equation}
\rm r_0(z)= r_0(0)(1+z) ^{-(3-\epsilon-\gamma)/\gamma} \label{eqn:r0z}
 \end{equation}
where r$_0(0)$ is the comoving correlation length at z$=$0, and $\epsilon$ gives the evolution of clustering (i.e., $\epsilon=0$ specifies a stable state, $\epsilon=\gamma-3$ describes no evolution, and $\epsilon=\gamma - 1$ corresponds to the linear growth of fluctuations). Throughout our analysis, we assume no evolution, or $\epsilon=\gamma-3$, which allows us to simplify Eqn. \ref{eqn:r0}. Over the relatively narrow redshift ranges for which we are integrating over, we can justify this assumption-- \cite{meneux} find that r$_0$ evolves by $\lesssim$ 20\% between z$=$0.2 and z$=$1.2 for late type galaxies and $\lesssim$ 50\% for early type galaxies. 

Auto-correlations compare distributions of a sample with itself, while cross-correlations compare distributions of one set of objects with a separate sample. Naturally, in the auto-correlation cases, $n_1= n_2$ in Eqn. \ref{eqn:r0}. We infer the real-space correlation length, r$_0$ and power-law slope, $\gamma$, by fitting the angular correlation function. 

Along with slope and amplitude parameters, we additionally calculate the Integral Constraint (IC) correction. In principle, our observed field may be superimposed on a void or overdensity, such that the observed number density in our field will not match the average number density in the universe. In theory, the integral in Eqn. \ref{eqn:wtheta} should average to 0 over all scales, i.e., since there are N galaxies in the universe, excess at small scales must be balanced by deficits at larger scales. We correct the observed angular correlation function by the IC factor defined in \cite{Roche99}.
 
\subsection{Measuring the angular correlation function}\label{sec:meas}
The angular correlation function, w($\theta$), measures the excess of clustering compared to a uniform, random distribution. To simulate the random distribution appropriately, we first create masks from the COSMOS+\GALEX fields. These masks mark the locations where bright, saturated stars and field edges prevent galaxies in those locations from appearing in the catalog. The black regions in Fig. \ref{fig:skydist} display masked locations. Populating the allowed, un-masked regions with random galaxies, we create the Random set. The histogram of separations in the Random set and Data sets are referred to as RR($\theta$) and D$_1$D$_2$($\theta$), respectively, where the subscripts 1 and 2 denote two separate data samples; D$_1$D$_2 =$DD in the auto-correlation case. Similarly, the separations between galaxies in the Random and Data sets are represented by D$_1$R and D$_2$R, or simply DR for autocorrelations. It has been shown that the Landy-Szalay \citep{LS} estimator is one of the most robust ways of measuring w($\theta$), reducing the edge effects of finite fields. Here we give the expression for the angular correlation function that we use to measure auto-correlations (Eqn. \ref{eqn:auto}) and cross-correlations (Eqn. \ref{eqn:cross}):
\begin{eqnarray}
{\rm w}(\theta) &=& {\rm \frac{r(r-1)}{n(n-1)}\frac{DD}{RR}-\frac{2 (r-1)}{n}\frac{DR}{RR}}+1 \label{eqn:auto} \\
 &= & {\rm \frac{r(r-1)}{n_1 n_2} \left (\frac{D_1D_2}{RR} -\frac{n_1}{r} \frac{D_2R}{RR}-\frac{n_2}{r}\frac{D_1R}{RR}+\frac{n_1 n_2}{r(r-1)}\frac{RR}{RR} \right )} \label{eqn:cross}
\end{eqnarray}

where r and n refer to the number of galaxies in the Random and Data samples (the subscripts 1 and 2 for n$_1$ and n$_2$ distinguish the separate samples, in the cross-correlation case). Our angular correlation function errors have been derived from boot-strapping, as described in \cite{BBS}. Cosmic variance adds another important source for error. However, unlike the bootstrap errors which are scale dependent, the errors from cosmic variance shift the entire correlation function. Using \cite{SPF} to calculate cosmic variance errors for our samples, we show these errors in our correlation function figures as shaded regions (for example, see Figs. \ref{fig:fit}-\ref{fig:acor_comp}). 

As described in $\S$\ref{sec:data}, our samples span the redshift range 0.4$<$z$<$1.2. While we may reduce statistical noise by choosing the largest sample possible -- selecting galaxies in this entire redshift span, correlating data across such a wide redshift range has disadvantages. By correlating data in two-dimensions, which aren't physically correlated in three-dimensional, de-projected space, we risk weakening the actual correlations. Therefore, we select two bin sizes: dz$=$0.2 (resulting in 4 overlapping bins: 0.4$-$0.6, 0.6$-$0.8, \ldots, 1.0$-$1.2) and dz$=$0.4 (2 bins: 0.4$-$0.8, and 0.8$-$1.2). Comparing the angular correlation functions for different redshift bins requires converting the angular correlation function into physical de-projected units, by the following reasoning:
\begin{eqnarray}
{\rm w(\theta)}&=& {\rm  \frac{\frac{dN(\Omega)}{d\Omega}-\left < \frac{dN}{d\Omega}\right > _0}{\left< \frac{dN}{d\Omega}\right> _0}} \label{eqn:wthetadef}\\ 
{\rm w_p(r_p)} &=& \int _{-\infty} ^{+\infty} \xi([{\rm r_p^2+\pi^2]^{1/2}) d}\pi \\
&=&{\rm  \frac{\frac{dN(r_p)}{dA}-\left < \frac{dN}{dA}\right >_0}{\left< \frac{dN}{dV}\right> _0}}
\end{eqnarray}
where r$_{\rm p}$ is the transverse distance and $\pi$ is the line-of-sight separation; $\left < \frac{\rm{dN}}{\rm{dA}} \right > _0$,  $\left < \frac{\rm{dN}}{\rm{d\Omega}} \right > _0$, and $\left< \frac{\rm dN}{\rm dV} \right>_0$ are, respectively, dN/dA, dN/d$\Omega$ and comoving number density for a uniform, non-clustered distribution. 
Finally, multiplying the w($\theta$) equation (Eqn. \ref{eqn:wthetadef}) by d$\Omega$/dA and averaging these expressions over the redshift bins, we derive
\begin{eqnarray}
{\rm \frac{w(\theta)}{w_p(r_p)}} &=& \frac{\left < {\rm\frac{dN}{dV}}\right >}{\left <{\rm \frac{dN}{d\Omega}} \cdot {\rm \frac{d\Omega}{dA}} \right > } \\
{\rm \frac{d\Omega}{dA} }&=& {\rm \frac{1}{D_A^2 (1+z)^2}} \\
{\rm w_p(r_p)} &=& \frac{\left < \frac{\rm dN}{\rm d\Omega}\cdot \frac{1}{\rm D_A^2 (1+z)^2} \right >}{\left < \frac{\rm dN}{\rm dV}  \right >} {\rm w(\theta)}
\end{eqnarray}
 where D$_{\rm A}$ is the angular diameter distance. As such, we can compare how the correlation function of the narrower bin sizes compare to the wider redshift ranges. Fig. \ref{fig:acor_all} shows how the wide bins differ from the narrow ones for samples with 0.8$<$z$<$1.2; we plot the ratio of the correlation functions for the wide to narrow bins, displaying both w($\theta$), top, and w$_{\rm p}$(r$_{\rm p}$), bottom. Since the discrepancy was not severe, we decided that the wider redshift bins would provide better statistics. Given the low abundance of UVLGs, their auto-correlation only proved possible for the full 0.4$<$z$<$1.2 sample; and for the LBAs,  their auto-correlation proved too noisy to be useful, and we were only able to attain reasonable results by cross-correlating with the largest sample: the ALL sample with dz$=$0.4 bins.  

For each sample, we calculate separate fits to the angular correlation function in two regimes: large scales ($>200$kpc) and small scales ($<200$kpc). We create a grid of a$_w$, spaced logarithmically, and delta, linearly spaced, values, with the best fit values corresponding to the minimum $\chi^2$. The errors on this fit are derived as the values corresponding to min$(\chi^2)+1$, and we make certain that the grid values extend significantly past these error values to avoid restricting the errors within the minimum or maximum value. The large scale is fit, additionally to the slope and amplitude, with the IC correction, discussed in the previous section. We show examples of our fitting to the separate samples in Fig. \ref{fig:fit}. 

 \subsection{Calculating Physical Quantities} \label{sec:quant}
The number density is required for calculating both Limber's Equation and the pair fraction. For our analysis, we use the observed number density. However, as a check we calculate the predicted physical number densities for the ALL, BLG, and LRG samples. We use $\alpha$, M$^*_{\rm B}$ (given in Vega magnitudes, and converted to AB magnitudes using \cite{Willmer}), and $\phi^*$ given for the COMBO-17 data in \cite{Faber} to determine the luminosity function, interpolating the LF between the specified redshift-dependent values to derive a smooth n(z) across our entire redshift range. We use a Gaussian kernel with $\sigma=0.02$, simulating the photo-z errors, to convolve the derived n(z)-- the results are shown as black dash-dotted lines in Fig. \ref{fig:zdist}, compared to the observed numbers. With the exception of some peaks and valleys, the observed BLG and LRG samples follow the predicted values well. The agreement between the predicted and observed ALL distribution is poorer-- at higher redshifts the observed number density appears to be underestimated. It is possible that the peaks in the observed n(z) in the LRG sample correspond to real structure. For the UVLG and LBA samples, we predict the number density using the calibrated luminosity function from \cite{Arnouts04}. In Table \ref{tab:LFfit}, we list the values used to fit the luminosity function to derive n(z) for all of our samples. For use in Limber's Equation, we smooth the observed redshift distribution, performing a boxcar average with a kernel size: $\Delta$z$=$0.08. The discrepancy between results derived using the observed versus predicted number densities is negligible.

Quantifying the smaller scales, we relate the fits from the inner regions (shown in red in Fig. \ref{fig:fit}) to pair fractions and merger rate, similar to analysis done by \cite{Bell06} for massive, luminous galaxies. The pair fraction is derived from integrating  Eqn. \ref{eqn:xi} from 0 to some maximum radius, within which galaxies are assumed to eventually merge-- the pair fraction quantifies the number of galaxies that are found within some critical, merging radius of another galaxy. However, since we are interested in the physical separation between the galaxies rather than the comoving distance, we replace the comoving correlation length, $\rm r_0$, with the physical correlation length, $\rm r_0/(1+z)$. Therefore, we calculate pair fractions by simplying the integral of Eqn. \ref{eqn:xi} with the following assumption: $\xi ({\rm r}<\rm {r_f}) >> 1$, into this expression
\begin{equation}
{\rm P(r<r_f)} = \frac{4\pi n_2}{3-\gamma} \left (\frac{{\rm r}_0}{1+z}\right)^\gamma(\rm {r_f})^{3-\gamma}.  \label{eqn:pf}
\end{equation}
Here, $n_2$ is the number density of the second sample in the cross-correlation. Following \cite{Bell06}, we also use r$_{\rm f}$=30 kpc to define close, physical pairs. Our errors for the pair fraction include cosmic variance uncertainty, estimated using the relations from \cite{SPF}. While referred to as a probability, the pair fraction is the mean number of galaxies found within r$_{\rm f}$ (See \cite{Patton00}) and may exceed unity. The pair fraction has quantitative power in comparing small scale clustering between different samples. 

The merger rate is derived by dividing the pair fraction by the timescale for this merging event to occur (assuming that the close pair merge into one merger remnant, such that the pair fraction is equal to twice the merger fraction). Though the calculation is uncomplicated, the steps for deriving the merger timescale are quite complex and fraught with uncertainties. \cite{KW} have compared pair fractions derived from their mock catalogs with merger rates from the Millenium Simulations to determine merger timescales. From these results, we infer that the merger timescale is $\sim 3000$ Myr h$^{-1}$, using the most general cross-correlation case of UVLGs with ALL and estimating $<$ M$_*>=10^{10}$ M$_\odot$ \citep{choopes} in the range z$=$0.4$-$1.2. Using this value, we determine merger rates for our data, shown in Table \ref{tab:pf}. According to \cite{KW}, merger timescales have been underestimated by $\sim 10$ in previous studies-- for example, \cite{Bell06} estimate the merger timescale to be 400 Myr. To avoid the large uncertainties in the merger timescale, we compare our pair fraction results to other studies, rather than comparing merger rates. 

\section{Results}\label{sec:results}

Given the fits shown in Fig. \ref{fig:fituvlg}, we derive the correlation strength (r$_0$) for UVLGs. We determine that UVLGs have r$_0=4.8 ^{+11.6}_{-2.4}$ h$^{-1}$Mpc (see Table \ref{tab:lsauto}), corresponding to M$_{\rm halo}>10^{11}$ (see \cite{MW}).  Figure \ref{fig:r0_uvlg} shows how our results compare with other published data, at various redshifts. The shaded region in this Figure corresponds to the shaded arrow in Figure 11 of \cite{adel05}-- this marks how r$_0$ evolves with redshift for a halo of mass 10$^{11.2}$\msun (bottom of shaded region) to 10$^{11.8}$\msun (top of shaded region). We discuss the evolution of UV-selected galaxies, in relation to other samples, in the following section.   

Studying the autocorrelations of the separate samples with each other, we find that typically (at all z) LRGs cluster most strongly with themselves at small scales, compared to auto-correlations of BLGs or ALL samples (top panel of Fig. \ref{fig:acor_comp}). Comparing auto-correlations of the samples with cross-correlations with UVLGs in the bottom panels of Fig. \ref{fig:acor_comp}, we find that at small scales (r$_{p}< 200$kpc), UVLGs cluster less strongly with LRGs compared to LRGs with themselves (\ie \cite{masjedi}); however, it is difficult to differentiate the UVLG cross correlations with the ALL and BLG samples compared to their auto-correlations. 

We compare the pair fractions of the autocorrelation of the other three samples with their cross-correlation with UVLGs in Fig. \ref{fig:comp}\footnote{While the pair fraction approaches and exceeds unity for the ALL auto- and cross-correlation results, this does not suggest that $\sim$100\% of the galaxies are found in pairs. Rather the pair fraction represents the mean number of secondary galaxies likely found near the primary ALL galaxy. See discussion in \S\ref{sec:quant}.}. At the smallest scales the ALL sample pairs more with UVLGs than with themselves, and given the large errors, the BLG sample galaxies appear to pair with UVLGs with equal likelihood to pairing with themselves. Yet, UVLGs are not likely companions of LRGs. Finally, the LBA-All (purple) pair fraction in Table \ref{tab:pf} has excessively large errors, preventing a meaningful comparison of the LBA-ALL sample with these other cases.

As we have discussed, small scale clustering translates to merger rates-- suggesting that UVLGs appear to merge often with ALL sample galaxies (and possibly, with other blue, star-forming galaxies), and do not merge often with LRGs. However, as there are some small scale pairs of UVLG-LRGs, these systems do interact on some occasions. Given that blue, star-forming galaxies typically do not reside in clusters, and UVLGs merge with these galaxies as often as they do with themselves, we expect that UVLGs will inhabit lower mass halos rather than massive clusters-- further constraining our results. 

Comparing with other results, we show our pair fraction results for the ALL, BLG, and LRG samples with other studies in Fig. \ref{fig:pfpub}. \cite{Conselice903} observe a sample of M$_{\rm B}<-18$ galaxies (orange squares) to measure merger fractions, assumed to be half of the pair fractions since two pair galaxies interact in a single merger; \cite{Lin04} study pair fractions for galaxies with $-22\le$M$_{\rm B} \le-20$ (green triangles), and the magenta star marks the result for  M$_{\rm B}<-20$, found by \cite{Bell06}. Since all of these samples are magnitude limited and color-selection independent, the ALL sample is most comparable (M$_{\rm g} < -18$). It should be noted that many studies (e.g. \cite{Lin04}, \cite{Conselice06}, \cite{Bell06}) find that pair fractions are extremely sensitive to the limiting magnitude. In order to compare our pair fraction results with other work, we adopt the observation from \cite{Bell06} that the correlation parameters are less sensitive to limiting depth than the number density. Therefore, we modify the derived pair fraction for our ALL sample, using Eqn. \ref{eqn:pf} with the small-scale correlation function parameters in Table \ref{tab:acor}, but changing the number density to represent M$_{\rm B} <-20$ selection. The modified pair fraction is shown as open circles in Fig. \ref{fig:pfpub} and agree well with published results. 

\section{Discussion}\label{sec:disc}

The picture that has emerged from this analysis depicts UVLGs as galaxies that inhabit halos with M$_{\rm halo}>10^{11}M_\odot$. UVLGs are more likely to interact and eventually merge with other galaxies (ALL) than the ALL sample does with itself, and they are as likely to interact and have companions as other blue, star-forming galaxies. These results are consistent with other studies showing that high SFRs may be closely related to increased merger rates. UVLGs are unlikely to interact with LRGs, which are expected to reside at the centers of massive clusters, providing another constraint regarding UVLG halo masses. In this section, we first place UVLGs amongst other categories of star-forming galaxies; then, we discuss our findings in light of other work related to UV-selected galaxies, star-formation and environments. 

\subsection{UVLGs vs. other star-forming galaxies}
 In Figure \ref{fig:r0_uvlg}, we display the large scale environmental measure for several other classes of star-forming galaxies. The BM, BX, LBGs \citep{adel05} are selected by color and limiting magnitude to be both luminous and blue, exhibiting the Lyman Break in their spectra. As discussed in the introduction, the LBAs have been found to have similar colors and specific star formation rates (SFR/M$_*$), suggesting similar star formation histories and stellar populations. In \cite{me2}, we found evidence for recent and continuous star formation activity, similar to the results found for z$=3$ LBGs by \cite{erb} and \cite{Shapley01}. We argue that LBAs are most comparable to LBGs. 
 
The FUV$<-18.25$ sample by \cite{Heinis} is a less extreme sample (extending to fainter magnitudes) of UV-selected galaxies compared to UVLGs. The other star-forming samples (BLGs, \cite{McCracken08}, \cite{Coil04}, \cite{Coil08}, \cite{gilli07}) are selected based on color and magnitude, and span a larger range of SFRs and stellar masses. As demonstrated in \cite{Heinis} for the local universe, the correlation strength appears to decrease with increasing UV luminosity. Yet, blue galaxies appear to have similar or lower r$_0$ than the \GALEX sample, including, the UVLG sample.

\cite{GK04} finds that around a critical stellar mass of $\sim 10^{10.3}$ M$_\odot$, galaxies separate into two classes: massive, red, with low SFRs versus less massive, blue, with high SFRs. The typical UVLG stellar mass is $\sim10^{10.5}$M$_\odot$ (LBAs have $10^{9.0}<$M$_*/\rm{M}_\odot<10^{10.7}$)\citep{choopes}. While extremely star-forming and blue, these galaxies have masses placing them on the boundary between these two classes-- suggesting, that they may be in the process of transitioning. This in-between stage is not well-understood as yet, but mergers and environmental processes may drive the transition in this regime. 

The more massive, infrared counterpart to UVLGs are Luminous Infrared Galaxies (LIRGs) and the more extreme Ultra-Luminous Infrared Galaxies (ULIRGs) -- dusty star-forming galaxies with significant emission coming from central 1kpc. They experience significant merger events in the local Universe \citep{sanders}. As yet, little research has been completed on studying the environments of local ULIRGs. According to \cite{zauderer}, mostly z$<0.3$ ULIRGs occupy the field environment, with few cases having cluster densities of Abell richness class 0 or 1. Studying LIRG pair fractions at z$\sim 1$, \cite{Bridge07} find that $\sim 27\%$ of the IR luminosity density arises from close pairs, while morphologically classified mergers contribute to $\sim 34\%$-- thereby, $\sim 61\%$ of the IR luminosity density can be attributed to early interactions (as evidenced by close pairs) and mergers (morphologically determined). While UVLGs are not dusty, it seems that UVLGs and LIRGs/ULIRGs share environmental conditions, possibly contributing to their high SFRs. 

\subsection{Effects at Large scales: Importance of Halo mass}
\cite{Conroy08-2} investigate the relationship between star formation, stellar mass and halo mass. Assuming that a tight relationship between stellar mass and halo mass exists, based on observational ``abundance matching'', they theoretically derive how mass assembly and star formation history evolve with time. Based on averaged results (they caution that exceptions to their results are completely plausible), \cite{Conroy08-2} determine that the relationship between star formation rate (SFR) and halo mass is Gaussian-shaped, whose mean and normalization parameters evolve slightly with redshift. However the broadness of the SFR peak, spanning 11.9$< {\rm log(M_{halo}/M_\odot)}<$13.1 for z$\sim$0, suggests that a distinct characteristic halo mass leading to higher SFRs does not exist. Rather galaxy properties likely shift gradually across the halo mass scale. 

As discussed earlier, \cite{Keres} suggest two regimes of gas fueling and star formation efficiency, depending on halo mass: hot-mode accretion for massive galaxies and halos masses $> 10^{11.4}$ M$_\odot$-- which relate to lower SFRs, and cold-mode accretion in less massive galaxies, where gas is drawn along filamentary structure and able to feed star formation more efficiently. Both of the results from \cite{Conroy08-2} and \cite{Keres} imply that the range of allowed halo masses for high SFRs could be broad-- with the possibility that there is some mixing of both hot- and cold- accretion modes for halos with masses, M$_{\rm halo} \sim 10^{12} M\odot$. 

While the study by \cite{Conroy08-2} has been very thorough for z$<$2, they do allow that the situation might differ at higher redshifts. As illustrated by \cite{Heinis}, at high redshifts observations show a strong correlation between rest-frame UV-luminosity and correlation length, or halo mass \citep{adel05, Arnouts02, foucaud, gia}. Yet, their results show that the nearly linear relation between UV-luminosity, or SFR, and halo mass seems to level off, or possibly even reverse at low redshifts \citep{Heinis} . 

Although we have studied a rare sample of galaxies, their close relation to well-studied high redshift LBGs allows us to make a direct comparison about the halo mass condition for LBG-like properties. \cite{adel05} find that LBG halo masses are $\sim 10^{12} {\rm M}_{\odot}$. The UVLGs have halo masses that are consistent with these values. While the errors are admittedly too large to be conclusive, our result hints that the halo masses in extreme star-forming galaxies, such as LBGs and UVLGs, are consistent with masses $\sim 10^{12}{\rm M}_\odot$. 

\subsection{Small Scale Clustering: Influence of mergers}
At large scales, we measured halos masses of UVLGs, in an attempt to connect star formation properties with large scale processes; at small scales another important environmental quantity is measured-- the likelihood for mergers. It has been shown by \cite{berrier} that halos do not necessarily merge in the same way that galaxies merge. At small scales, the more relevant quantity is the Halo Occupation Distribution (HOD), or the way in which galaxies occupy their halos. This information constrains galaxy formation models at the $\le$ 100 kpc scales. 

Studying their small scale clustering, we find that UVLGs and LBAs tend to have high pair fractions and merger rates with the ALL sample. \HST images of the LBAs offer similar clues, displaying companions or ongoing interactions (such as tidal features and plumes) in the high resolution images \citep{rod}. Consistent with the results found by \cite{Li08}, the most extreme star-forming galaxies are preferentially found in close pairs. Studying LBGs at high redshifts (z$\sim 4-5$), \cite{Lee06} also find evidence for small-scale clustering. They interpret that LBGs likely have fainter companions occupying the same halos, and find that the HOD model fits their data well. 

Using merger histories to connect z$=3$ LBGs to present-day massive galaxies, \cite{Conselice06} argue that M$_*>10^{10}~\rm M_\odot$ galaxies undergo $\sim$ 4 major mergers, with most of the interactions having occured before z$=$1.5. Furthermore, they show that z$=$0.5 galaxy with stellar mass $\sim10^{10}~\rm M_\odot$ likely originated as a 10$^{9}~\rm{M}_\odot$ galaxy, whose stellar mass has grown both by mergers and increased star formation activity. While the number of mergers plateaus for $10^{10}~\rm M_\odot$ galaxies by z$\sim$2, less massive galaxies seem to continue merging until z$=$0.5. UVLGs have stellar masses $10^{10.5}-10^{11.3}$M$_\odot$, and LBAs have lower masses ($10^{9.5} - 10^{10.5}$ M$_\odot$). If UVLGs follow the same merger histories and evolution as more general galaxies studied by \cite{Conselice06}, the present UVLG and LBA masses suggest that their z$=3$ predecessors had stellar masses between $10^{8}-10^{9}$ M$_\odot$ and are probably still actively merging at z$<1$. 

Could mergers be the main triggers for the increased star formation in these low dust, high star-forming galaxies? Our results, and more generally, the conclusion reached by \cite{Li08} find that interactions and mergers do lead to enhanced star formation. As discussed in the Introduction, this concept is not new -- having theoretical roots from \cite{Toomre2}, mergers efficiently funnel gas into central regions of galaxies, providing fodder for star formation. This theory is supported by several observational studies (see Reviews by \cite{keel} and \cite{Struck}). 

We can also reverse the query: do all mergers and interactions create UVLGs or LBGs?  Mergers and interactions do enhance star formation, but these processes are also seen in LIRGs and ULIRGs -- which are dustier systems, apparently different from UVLGs and LBAs. The overlap between these IR star-forming populations and UV-star-forming galaxies has not been studied in detail. We speculate that both of these systems may be related, as different phases of the merger process-- with various unknown factors such as orientation, timescales, and kinematics adding complexity to the evolution and final merger state; or, while some fraction of mergers result in UVLGs, some other fraction result in dustier LIRGs or a separate class of galaxies, altogether. \cite{Bridge07} has found that a larger fraction of LIRGs appear in the merger phase than in the pre-merger, close pair phase. So it is plausible that UVLGs relate to the earliest interaction phase, exhibiting high SFRs, low dust and distorted features. The highest restframe-UV surface brightnesses, observed in LBAs, corresponds to galaxies found to have a higher incidence of pairs. Unfortunately, further investigation about their environment is beyond the scope of this statistical study because of their rarity. Extrapolating from our data on local LBG analogs, it seems that close pairs are one relevant ingredient to their condition. However, it is unlikely that every merger will result in a low dust, high surface brightness, high SFR galaxy -- other (unknown as yet) constraints must also be placed in order to produce this particular breed of galaxy. 

\section{Conclusions}\label{sec:conc}
 
Using the COSMOS data to compare angular cross-correlation functions of UVLGs with three other samples: ALL, BLGs, and LRGs, we determine both large and small scale environments of UVLGs. In our analysis, we assume the following:
1.) The correlation function does not evolve with redshift. The full correlation function includes a term from evolution (See Eqns. \ref{eqn:xiform} and \ref{eqn:r0z}). We assume that $\epsilon=\gamma-3$, which describes non-evolution in the correlation function. This assumption is justified by the study by \cite{meneux} -- finding that r$_0$ evolves by $\lesssim$ 20\% over z$=$0.2$-$1.2 range. 2.) The correlation function can be characterized by a power law at large scales, and a separate power law at small scales (see \cite{ross}). 3.)The size of the redshift bins do not affect correlation results. Bins that are too wide might make real-space correlations appear weaker by projecting across an excessively wide redshift range. However, we assume that dz$=$0.4 is not too wide, and check this assumption by comparing to more narrow dz$=$0.2 bins-- finding that there is good agreement between the wide and narrow cases. 

We find that UVLGs inhabit halos with masses exceeding $10^{11}$M$_\odot$. This result is further constrained by the additional observation that UVLGs and LRGs (which typically trace the most massive halos) do not often form pairs. Rather, UVLGs form close pairs with higher probability compared to the ALL sample forming pairs with themselves; and UVLGs pair with BLGs with the same probability as BLG-BLG pairs. Therefore we conclude that mergers are relevant environmental effects responsible for triggering high levels of SFR, observed in UVLGs and the high redshift LBGs. We are not able to definitively measure the halo mass of this sample given our large errors. However, we explore the possibility that halo mass may play some role in the UVLG or LBG condition-- since both examples of UV-selected, high SFR galaxies have halo masses consistent with $\sim 10^{12}$M$_\odot$. 

\url[]{GALEX} (Galaxy Evolution Explorer) is a NASA Small Explorer, launched in April 2003. We gratefully acknowledge NASA's support for construction, operation, and science analysis for the GALEX mission, developed in cooperation with the Centre National d'Etudes Spatiales of France and the Korean Ministry of Science and Technology.  The HST COSMOS program was supported through NASA grant HST-GO-09822. More information on the COSMOS survey is available at http://www.astro.caltech.edu/cosmos. We thank Michael Blanton for access to the IDL kcorrect (version 4.1.4) analysis package. This work has greatly benefitted from the careful comments and suggestions made by the anonymous referee. A.R.B. gratefully recognizes Ian McGreer and Andrei Mesinger for their contributions to this analysis, and David Hogg for insightful discussions.

\begin{deluxetable}{ccc}
\tablecolumns{3} 
\tablewidth{0pc} 
\tablecaption{Sample Numbers} 
\tablehead{ 
\colhead{Sample} & \colhead{z} & \colhead{N}
}
\startdata 
    All  &   0.4 $-$ 0.8  &       60872   \\
    All  &   0.8 $-$ 1.2  &      107473   \\
\hline
    All  &   0.4 $-$ 1.2  &      168345   \\
\hline
    BLG  &   0.4 $-$ 0.8  &        7678   \\
    BLG  &   0.8 $-$ 1.2  &       25100   \\
\hline
    BLG  &   0.4 $-$ 1.2  &       32778   \\
\hline
    LRG  &   0.4 $-$ 0.8  &       11175   \\
    LRG  &   0.8 $-$ 1.2  &       15861   \\
\hline
    LRG  &   0.4 $-$ 1.2  &       27036   \\
\hline
   UVLG  &   0.4 $-$ 0.8  &         752   \\
   UVLG  &   0.8 $-$ 1.2  &        2737   \\
\hline
   UVLG  &   0.4 $-$ 1.2  &        3489   \\
\hline
    LBA  &   0.4 $-$ 0.8  &          50   \\
    LBA  &   0.8 $-$ 1.2  &         232   \\
\hline
    LBA  &   0.4 $-$ 1.2  &         282   \\
\hline
\enddata 
\end{deluxetable} 

 \begin{deluxetable}{cccccc}
\tablecolumns{6} 
\tablewidth{0pc} 
\tablecaption{Fits for Luminosity Function Used to Derive n(z)} 
\tablehead{ 
\colhead{Sample} & \colhead{$z$} & \colhead{$\alpha$} & \colhead{M$^*_{\rm B}$} & \colhead{$\phi^*$}  & \colhead{Reference}\\
\colhead{} & \colhead{} & \colhead{} & \colhead{(AB mag)\tablenotemark{a}} & \colhead{(10$^{-4}$ Gal Mpc$^{-3}$)} & \colhead{}
}
\startdata 
    All  &  0.3 &   -1.3  &   -21.10  &  32.26  &  \cite{Faber} \\
  All  &  0.5 &   -1.3  &   -21.30  &  33.32  &   \cite{Faber} \\
  All  &  0.7 &   -1.3  &   -21.62  &  32.16  &   \cite{Faber} \\
  All  &  0.9 &   -1.3  &   -21.35  &  41.26  &   \cite{Faber} \\
  All  &  1.1 &   -1.3  &   -21.48  &  34.72  &  \cite{Faber} \\
\hline
BLG   &  0.3 &   -1.3  &   -20.84  &  24.26  &    \cite{Faber} \\
BLG   &  0.5 &   -1.3  &   -21.20  &  23.20  &    \cite{Faber} \\
BLG   &  0.7 &   -1.3  &   -21.40  &  27.27  &    \cite{Faber} \\
BLG   &  0.9 &   -1.3  &   -21.20  &  37.32  &    \cite{Faber} \\
BLG   &  1.1 &   -1.3  &   -21.35  &  29.19  &    \cite{Faber} \\
\hline
LRG   &  0.3 &   -0.5  &   -20.73  &  21.91  &      \cite{Faber} \\
LRG   &  0.5 &   -0.5  &   -20.87  &  19.97  &      \cite{Faber} \\
LRG   &  0.7 &   -0.5  &   -21.20  &  17.75  &      \cite{Faber} \\
LRG   &  0.9 &   -0.5  &   -21.28  &  11.89  &      \cite{Faber} \\
LRG   &  1.1 &   -0.5  &   -21.68  &   5.32  &      \cite{Faber} \\
\hline
UVLG   &  0.3 &   -1.5  &   -19.53  &   6.00  &  \cite{Arnouts04} \\
UVLG   &  0.6 &   -1.5  &   -19.63  &  10.30  &   \cite{Arnouts04} \\
UVLG   &  1.0 &   -1.5  &   -19.89  &   8.00  &   \cite{Arnouts04} \\
\enddata 
\tablenotetext{a}{M$^*_{\rm B}$ values from \cite{Faber} have been converted from Vega magnitudes to AB magnitudes using \cite{Willmer}. }
\label{tab:LFfit}
\end{deluxetable} 

\begin{deluxetable}{ccccc}
\tablecolumns{5} 
\tablewidth{0pc} 
\tablecaption{Large-scale Auto-correlation Results} 
\tablehead{ 
\colhead{Sample} & \colhead{z}  & \colhead{A$_{w,\theta}$} & \colhead{$\delta$} & \colhead{r$_0$} \\
 &  & (10$^{-3} (^\circ)^\delta$) & &  \colhead{(h$^{-1}$Mpc)}
}
\startdata 
                 ALL    &      0.4 $-$ 0.8    &      5.0    $\pm{0.9}$  &     0.64      $^{+0.05} _{- 0.04}$  &     3.3       $\pm{0.4}$   \\
                 ALL    &      0.8 $-$ 1.2    &     22.0        $^{+4.2} _{- 3.8}$  &     0.32      $^{+0.04} _{- 0.03}$  &     3.8       $\pm{0.4}$   \\
\hline
                 BLG    &      0.4 $-$ 0.8    &     16.6       $^{+21.5} _{- 7.5}$  &     0.44      $^{+0.13} _{- 0.17}$  &     3.9      $^{+2.9} _{- 1.3}$   \\
                 BLG    &      0.8 $-$ 1.2    &     50.0      $^{+21.4} _{- 18.3}$  &     0.22      $^{+0.08} _{- 0.05}$  &     4.4      $^{+0.9} _{- 1.0}$   \\
\hline
                 LRG    &      0.4 $-$ 0.8    &      2.7        $^{+1.9} _{- 1.1}$  &     0.83      $^{+0.14} _{- 0.13}$  &     2.9      $^{+1.2} _{- 0.8}$   \\
                 LRG    &      0.8 $-$ 1.2    &     22.3        $^{+6.0} _{- 7.4}$  &     0.43      $^{+0.09} _{- 0.05}$  &     4.8      $^{+0.9} _{- 1.2}$   \\
\hline
                UVLG    &      0.4 $-$ 1.2    &     32.8     $^{+541.0} _{- 21.1}$  &     0.32      $^{+0.22} _{- 0.29}$  &     4.8     $^{+11.6} _{- 2.4}$   \\

                \enddata
  \label{tab:lsauto}
\end{deluxetable} 

\begin{deluxetable}{cccccccc}
\rotate
\tablecolumns{8} 
\tablewidth{0pc} 
\tablecaption{Small scale UVLG Cross-correlation Results} 
\tablehead{ 
\colhead{Sample} & \colhead{z}  & \colhead{A$_{w,\theta}$} & \colhead{$\delta$} & \colhead{r$_0$\tablenotemark{a}} & \colhead{n$_{2nd}$} &\colhead{f$_{\rm pair}$} & \colhead{R$_{\rm merge}$}\\
 &  & (10$^{-3} (^\circ)^{\delta}$) & &  \colhead{(h$^{-1}$Mpc)} & \colhead{(Mpc$^{-3}$)} & &\colhead{(Gyr$^{-1}$)}
}
\startdata 
          UVLG x ALL    &      0.4 $-$ 0.8    &     0.06        $^{+0.16} _{- 0.04}$  &      1.5    $\pm{0.2}$  &     1.3      $^{+1.1} _{- 0.5}$  &       0.084  $\pm{ 0.006}$  &        0.47      $^{+0.12} _{- 0.08}$  &       0.109      $^{+0.027} _{- 0.019 }$ \\ 
          UVLG x ALL    &      0.8 $-$ 1.2    &     0.03        $^{+0.03} _{- 0.02}$  &      1.5        $^{+0.1} _{- 0.1}$  &     1.3      $^{+0.6} _{- 0.4}$  &       0.159  $\pm{ 0.027}$  &        1.18      $^{+0.34} _{- 0.28}$  &       0.276      $^{+0.079} _{- 0.064 }$ \\ 
\hline
          UVLG x BLG    &      0.4 $-$ 0.8    &     0.07        $^{+0.92} _{- 0.07}$  &      1.5        $^{+0.5} _{- 0.4}$  &     1.4      $^{+3.8} _{- 1.0}$  &       0.011  $\pm{ 0.001}$  &        0.07      $^{+0.13} _{- 0.03}$  &       0.015      $^{+0.031} _{- 0.006 }$ \\ 
          UVLG x BLG    &      0.8 $-$ 1.2    &     0.24        $^{+0.59} _{- 0.17}$  &      1.2    $\pm{0.2}$  &     1.9      $^{+1.8} _{- 0.8}$  &       0.039  $\pm{ 0.012}$  &        0.14      $^{+0.08} _{- 0.06}$  &       0.032      $^{+0.019} _{- 0.013 }$ \\ 
\hline
          UVLG x LRG    &      0.4 $-$ 0.8    &     0.21       $^{+32.44} _{- 0.20}$  &      1.2        $^{+0.8} _{- 0.9}$  &     1.6     $^{+59.7} _{- 1.3}$  &       0.015  $\pm{ 0.001}$  &        0.03      $^{+0.07} _{- 0.01}$  &       0.007      $^{+0.017} _{- 0.003 }$ \\ 
          UVLG x LRG    &      0.8 $-$ 1.2    &     0.16        $^{+0.82} _{- 0.14}$  &      1.2        $^{+0.3} _{- 0.3}$  &     1.8      $^{+3.0} _{- 1.0}$  &       0.023  $\pm{ 0.010}$  &        0.11      $^{+0.09} _{- 0.06}$  &       0.025      $^{+0.022} _{- 0.014 }$ \\ 
\hline
           LBA x ALL    &      0.4 $-$ 0.8    &     0.22       $^{+10.41} _{- 0.21}$  &      1.3        $^{+0.6} _{- 0.7}$  &     1.9     $^{+21.0} _{- 1.5}$  &       0.084  $\pm{ 0.020}$  &        0.65      $^{+1.04} _{- 0.32}$  &       0.153      $^{+0.242} _{- 0.074 }$ \\ 
           LBA x ALL    &      0.8 $-$ 1.2    &     0.01        $^{+0.19} _{- 0.01}$  &      1.6        $^{+0.6} _{- 0.4}$  &     1.2      $^{+3.1} _{- 0.9}$  &       0.159  $\pm{ 0.060}$  &        2.61    $^{+52.64} _{- 25.78}$  &       0.608     $^{+12.283} _{- 6.016 }$ \\ 
                      \enddata 
 \label{tab:pf}
\end{deluxetable} 

\hspace{-0.5in}
 \begin{deluxetable}{cccccccc}
 \rotate
\tablecolumns{8} 
\tablewidth{0pc} 
\tablecaption{Small scale Auto-correlation Results} 
\tablehead{ 
\colhead{Sample} & \colhead{z}  &\colhead{A$_{w, \theta}$} & \colhead{$\delta$} & \colhead{r$_0$\tablenotemark{a}} & \colhead{n} &\colhead{f$_{\rm pair}$} & \colhead{R$_{\rm merge}$}\\
 &  & (10$^{-3}(^\circ)^{\delta}$) & &  \colhead{(h$^{-1}$Mpc)} & \colhead{(Mpc$^{-3}$)} & &\colhead{(Gyr$^{-1}$)}
 }
\startdata
                 ALL    &      0.4 $-$ 0.8    &     0.56        $^{+0.19} _{- 0.13}$  &     1.05      $^{+0.04} _{- 0.05}$  &     2.0      $^{+0.3} _{- 0.3}$  &       0.084  $\pm{ 0.015}$  &        0.14   $\pm{0.03}$  &       0.033      $^{+0.008} _{- 0.007 }$  \\
                 ALL    &      0.8 $-$ 1.2    &     0.30   $\pm{0.05}$  &     1.12      $^{+0.03} _{- 0.03}$  &     1.9    $\pm{0.2}$  &       0.159  $\pm{ 0.027}$  &        0.26   $\pm{0.05}$  &       0.060      $^{+0.012} _{- 0.012 }$  \\
\hline
                 BLG    &      0.4 $-$ 0.8    &     0.44        $^{+1.44} _{- 0.35}$  &     1.14   $\pm{0.24}$  &     1.9      $^{+2.4} _{- 1.0}$  &       0.011  $\pm{ 0.002}$  &        0.03   $\pm{0.01}$  &       0.007      $^{+0.003} _{- 0.003 }$  \\
                 BLG    &      0.8 $-$ 1.2    &     0.34        $^{+0.34} _{- 0.16}$  &     1.11      $^{+0.10} _{- 0.11}$  &     1.9      $^{+0.9} _{- 0.5}$  &       0.039  $\pm{ 0.010}$  &        0.07   $\pm{0.03}$  &       0.017      $^{+0.006} _{- 0.006 }$  \\
\hline
                 LRG    &      0.4 $-$ 0.8    &     0.38        $^{+0.52} _{- 0.23}$  &     1.23      $^{+0.15} _{- 0.14}$  &     2.0      $^{+1.2} _{- 0.7}$  &       0.015  $\pm{ 0.002}$  &        0.06   $\pm{0.02}$  &       0.015      $^{+0.004} _{- 0.004 }$  \\
                 LRG    &      0.8 $-$ 1.2    &     0.18        $^{+0.14} _{- 0.07}$  &     1.41      $^{+0.08} _{- 0.09}$  &     2.1      $^{+0.7} _{- 0.4}$  &       0.023  $\pm{ 0.004}$  &        0.24   $\pm{0.06}$  &       0.056      $^{+0.014} _{- 0.012 }$  \\
\enddata 
 \label{tab:acor}
\end{deluxetable} 

\begin{figure}[h]
   \centering
   \includegraphics[width=2.6in]{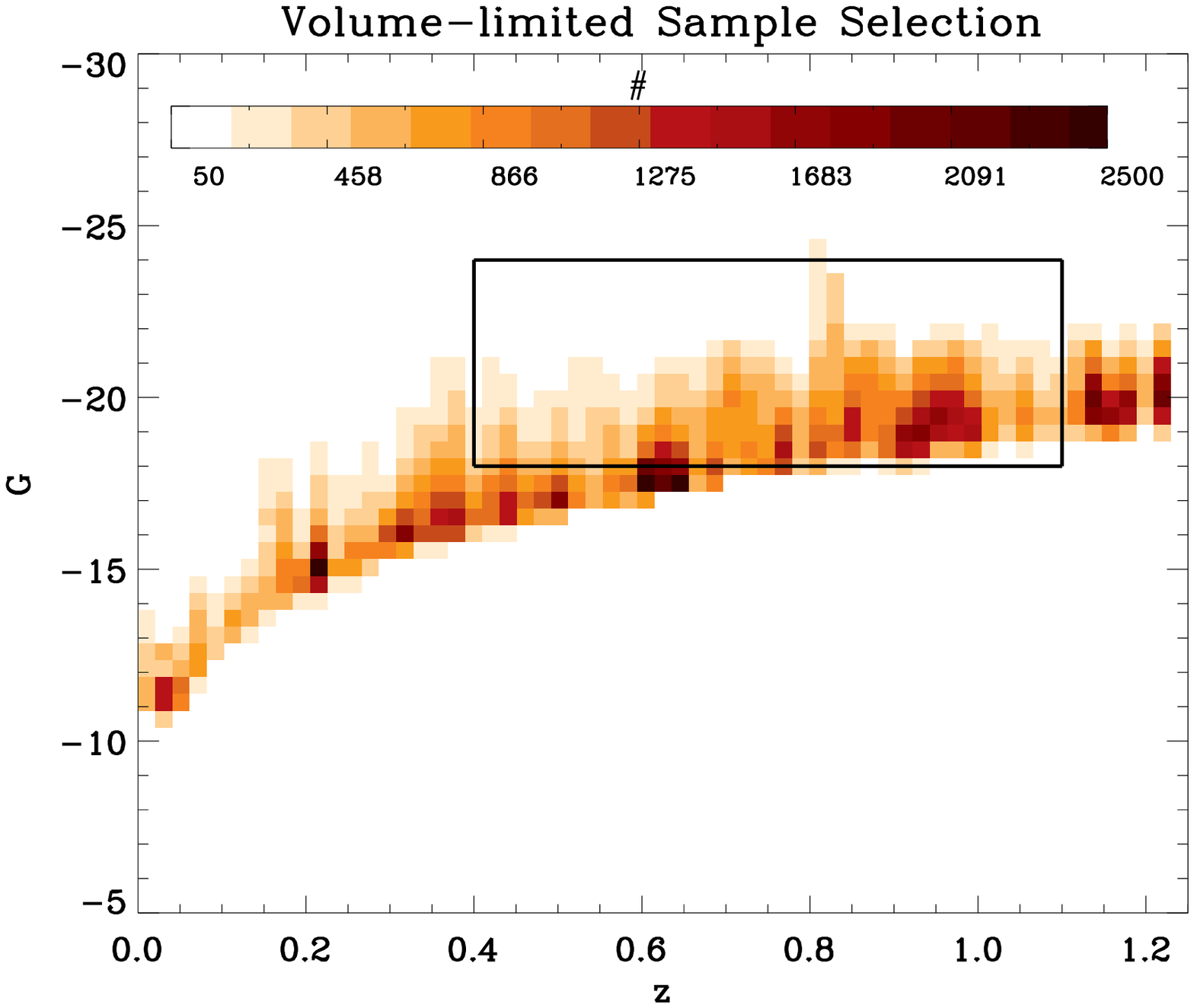} % requires the graphicx package
   \includegraphics[width=2.6in]{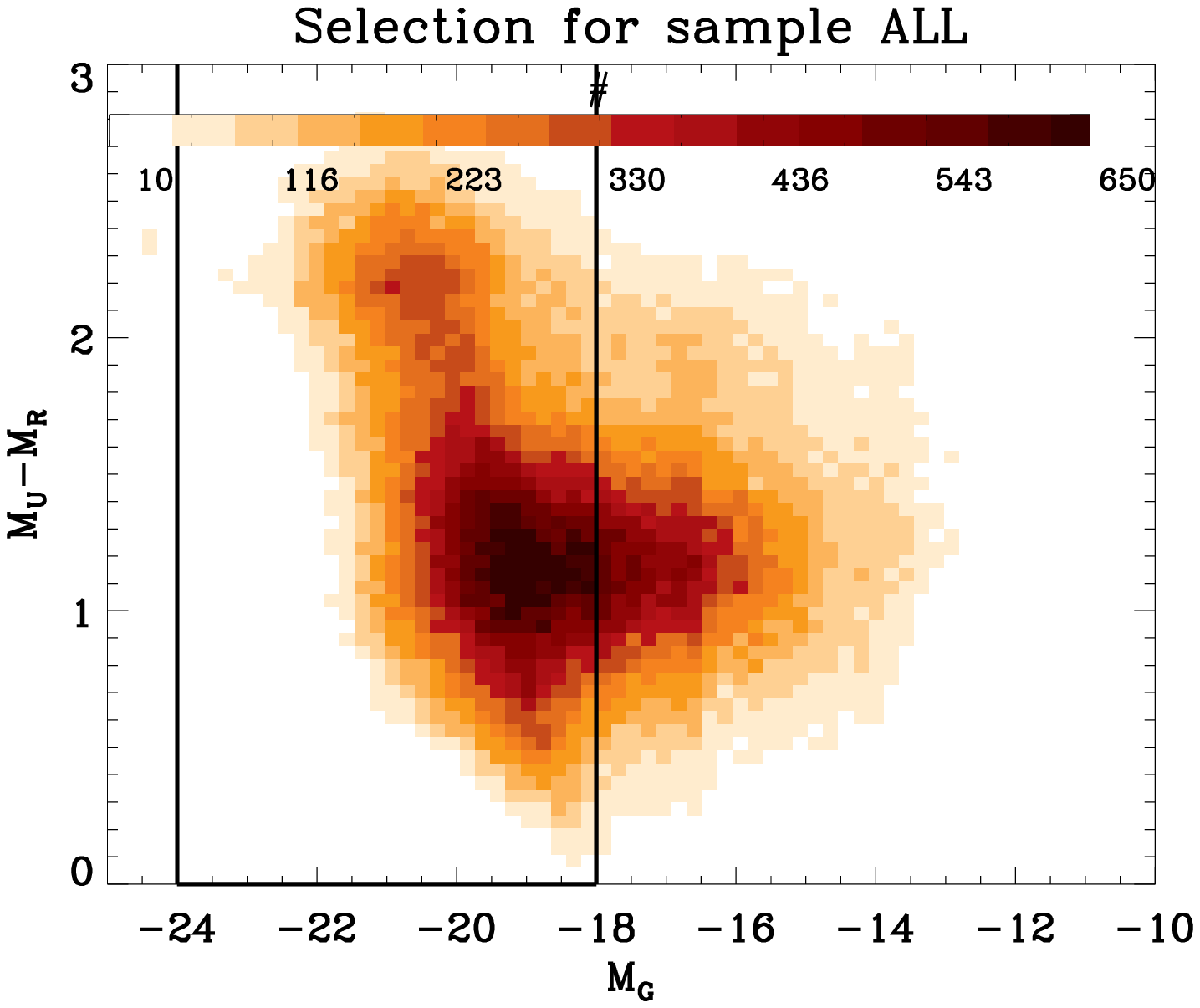} % requires the graphicx package
   \includegraphics[width=2.6in]{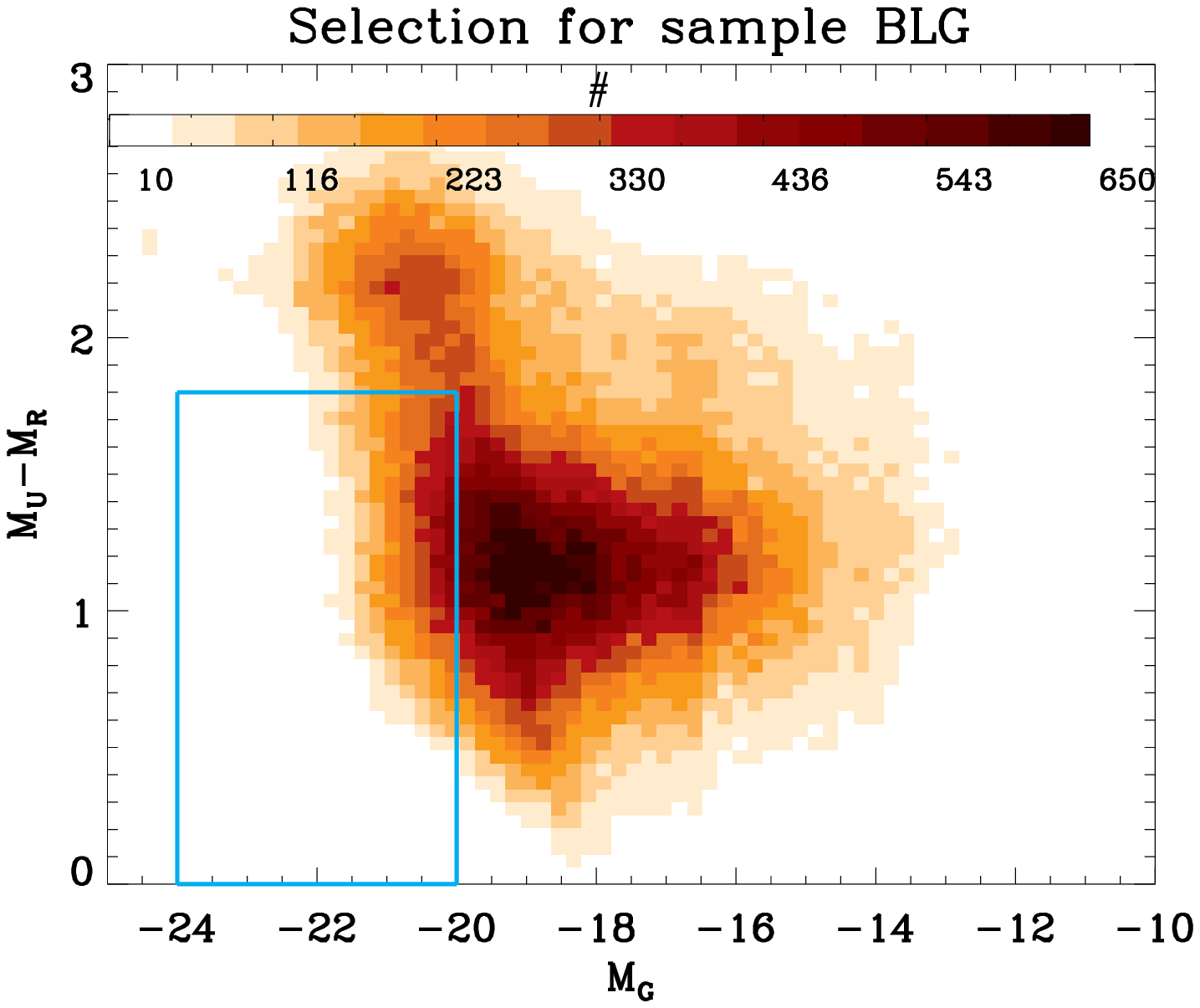} % requires the graphicx package
   \includegraphics[width=2.6in]{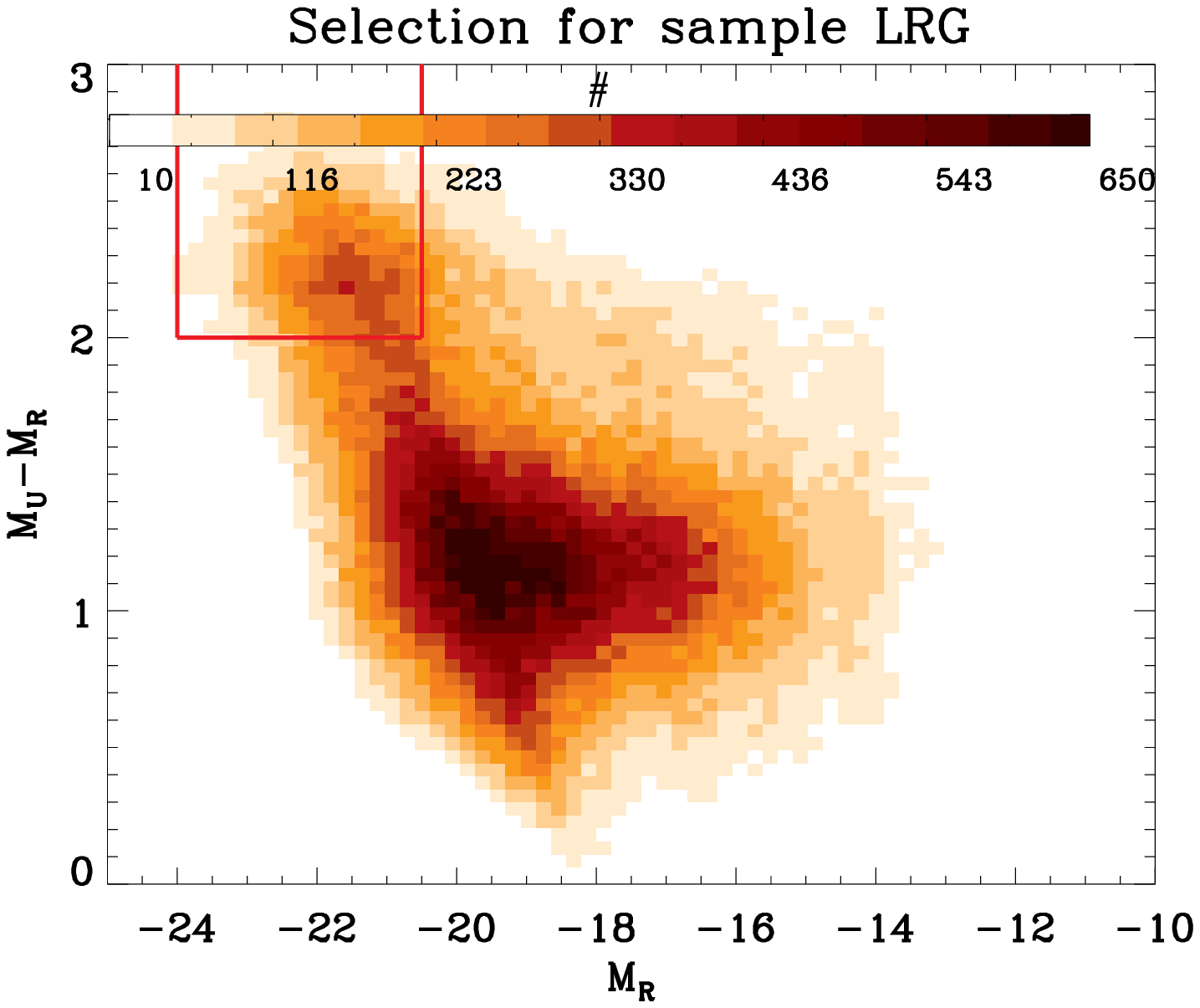} % requires the graphicx package
   \includegraphics[width=2.6in]{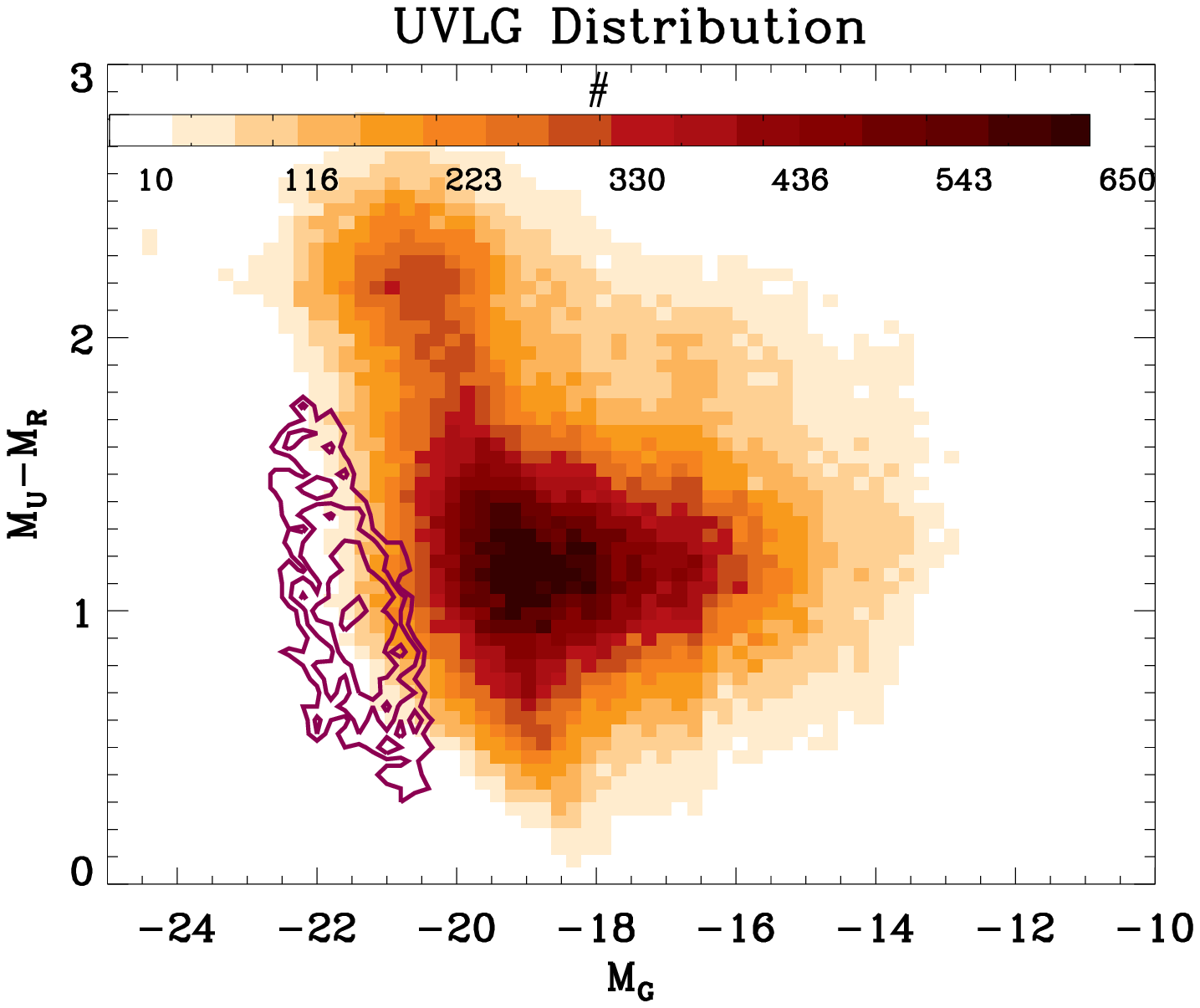} % requires the graphicx package
   \includegraphics[width=2.6in]{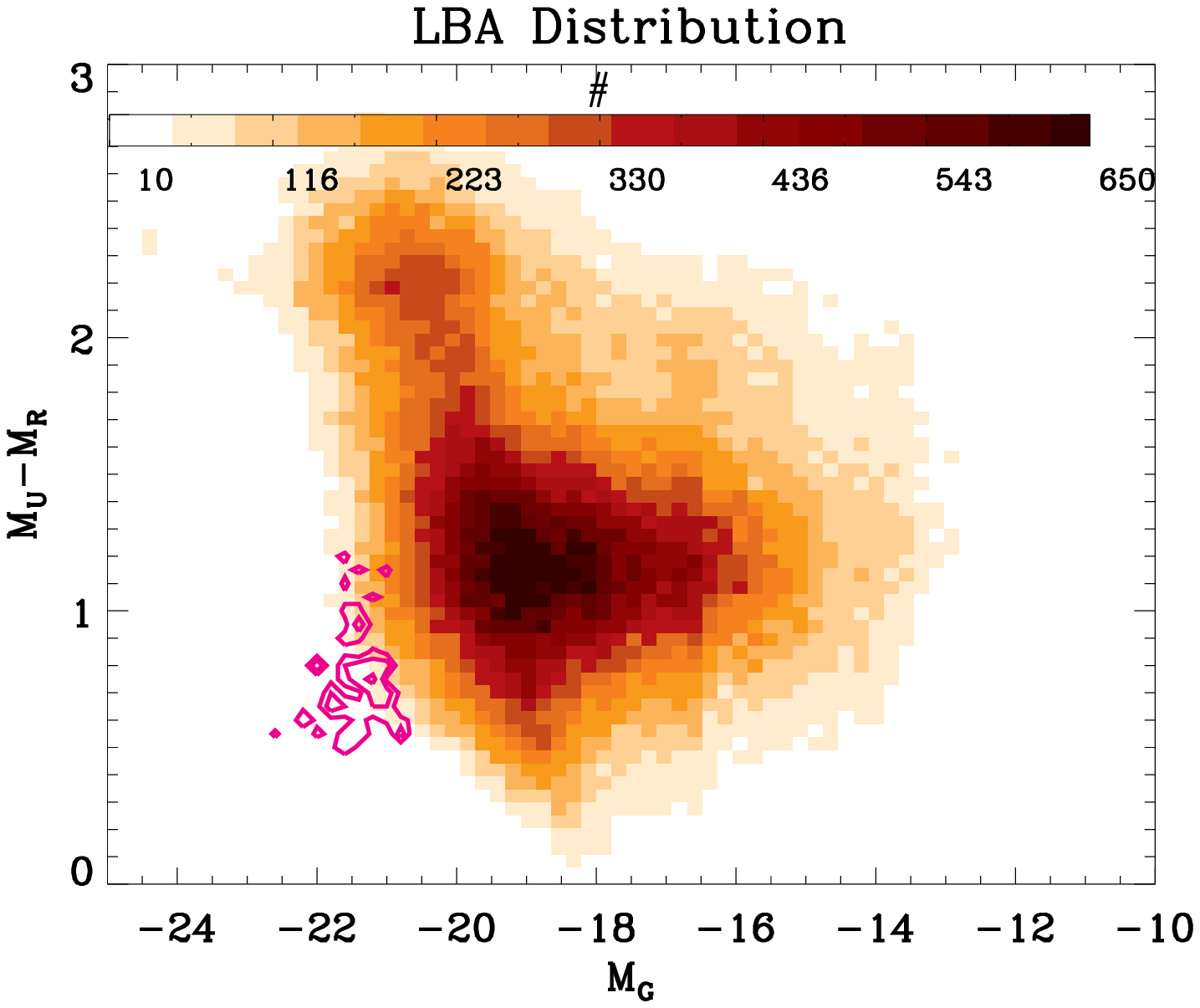} % requires the graphicx package
   \caption{Sample selection for the three comparison samples, drawn from the I-band selected (I$<$25.5) COSMOS catalog: ALL (upper panels), BLG (middle left), and LRG (middle right) and the color-magnitude distributions of UVLGs (lower left) and LBAs (lower right), selected from the COSMOS+\GALEX (I$<$25.5 and NUV$<$23) combined catalog. The upper left panel illustrates the luminosity-redshift distribution for the full data, with the black box defining a volume-limited sample (ALL). The plot on the upper right shows how this sample of galaxies, ALL, occupies the color-magnitude diagram. The lower plots show color magnitude diagrams for the other samples. We define BLGs to have $M_{\rm u}-M_{\rm r} <$1.8 and $M_{\rm g}<$-20.0 (shown on middle left) and LRGs to have $M_{\rm u}-M_{\rm r} >$2.0 and $M_{\rm r}<$-20.5 (middle right). The lowest left panel shows the UVLG distribution in purple with contour levels representing the number of galaxies in color-magnitide bins corresponding to (3, 6, 12, 24) galaxies; similarly, the lowest right panel shows the same for LBAs with contour levels corresponding to (1.5, 3, 6) galaxies. } 
   \label{fig:colmag}
\end{figure}
\begin{figure}[h]
   \centering 
     \includegraphics[width=7in]{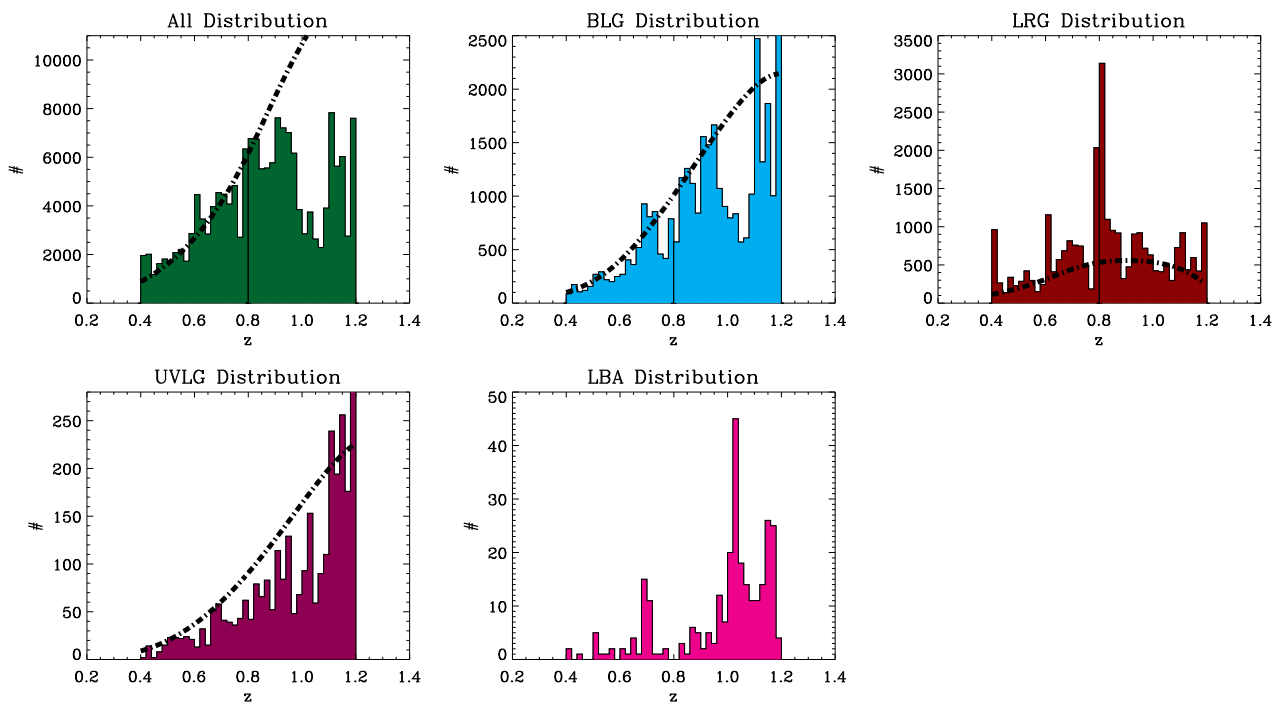} % requires the graphicx package
   \caption{Redshift distribution of all the separate samples -- ALL (top left), BLG (top middle), LRG (top right), UVLG (bottom left), and LBA (bottom right). For each sample, we divide the full redshift range into dz$=$0.4 bins: 0.4$-$0.8, and 0.8$-$1.2. Black dash-dotted lines show predicted number densities calculated from luminosity functions, given by \cite{Faber} (and \cite{Arnouts04}, for UVLGs). }
   \label{fig:zdist}
\end{figure}

\begin{figure}[h]
   \centering
   \includegraphics[height=6.5in]{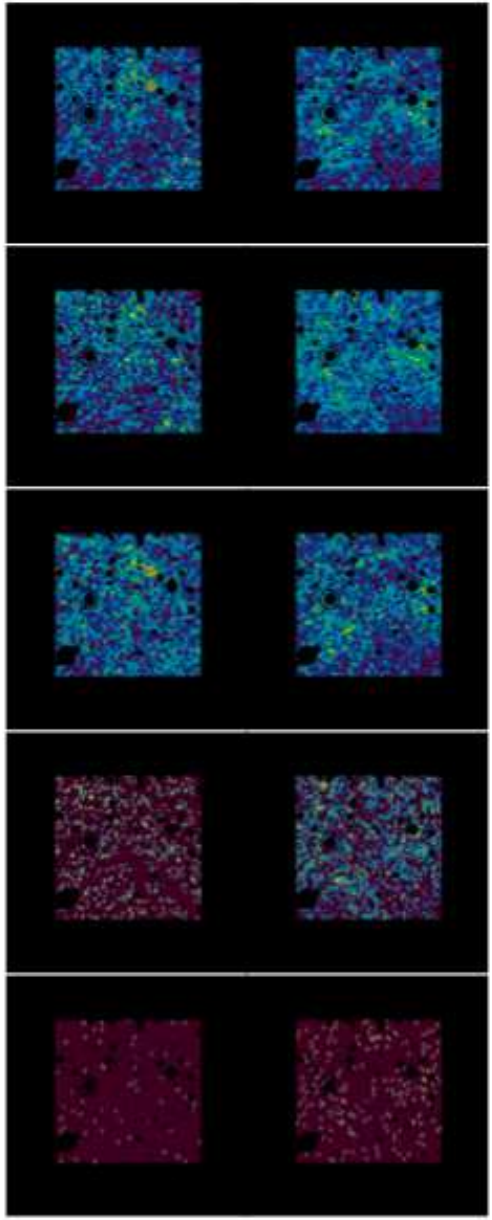} \\% requires the graphicx package

   \caption{Spatial sky distributions--0.4$<$z$<$0.8 (left), and 0.8$<$z$<$1.2(right) for samples from top to bottom: ALL (top), BLG, LRG, UVLG, and LBA (bottom). The distribution is smoothed to a scale of 4.8\arcmin. The spatial distributions are displayed over the mask (black denotes regions where no reliable observations exist)-- the allowed regions of the masks, appearing in these plots as dark violet regions underlying the colored points, are used for populating the random catalogs. Colors in this figure progress from deep violet to red with increasing density. }
   \label{fig:skydist}
\end{figure}

\begin{figure}[h]
   \centering
   \includegraphics[width=6.5in]{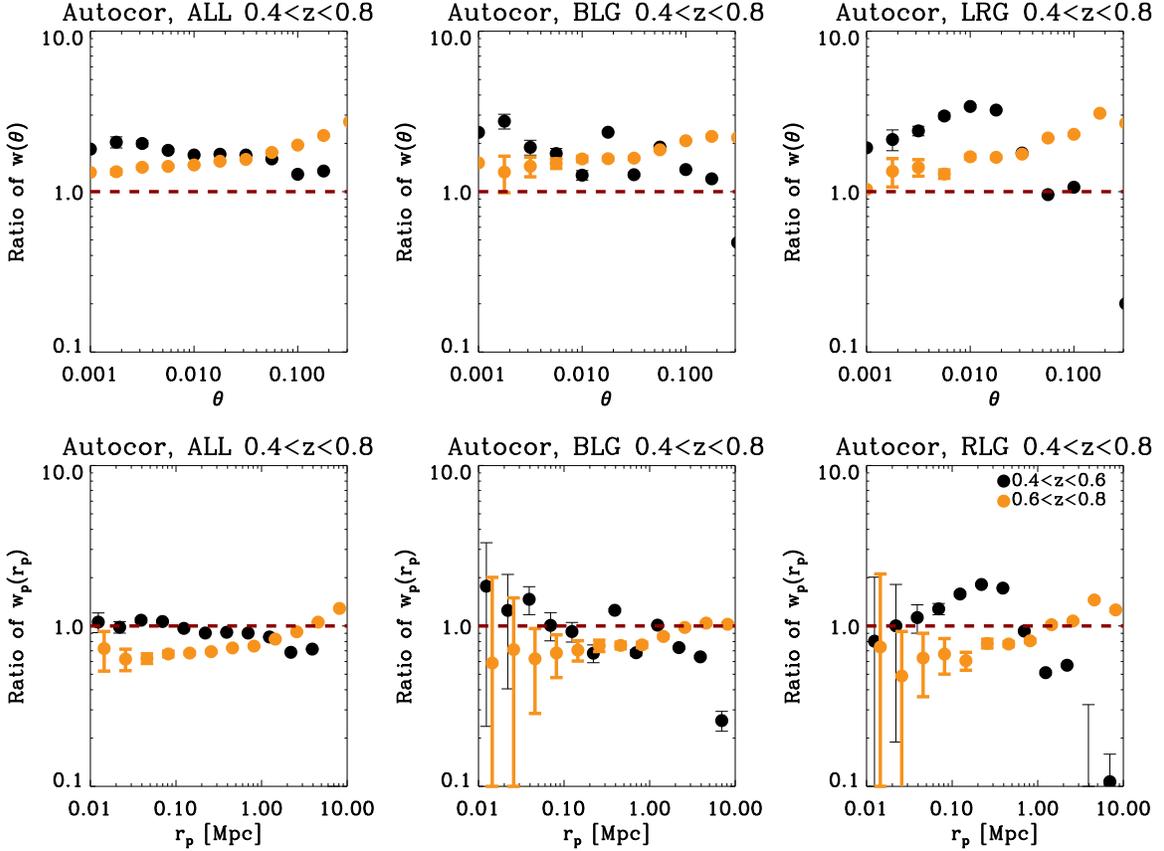} % requires the graphicx package
   \caption{We compare the auto-correlation results for the narrow (dz$=$0.2) to wide (dz$=$0.4) redshift bins, by displaying the ratios of w($\theta$), at top, and w$_p$(r$_p$), at bottom. For simplicity, we have chosen to display only one representative redshift range (0.4$-$0.8) for the autocorrelation of the ALL (left), BLG (middle) and LRG (right) samples. The ratios of the 0.4$<$z$<$0.6 data to the wider 0.5$<$z$<$0.9 are shown as black points, and the ratios of 0.6$-$0.8 to 0.5$-$0.9 appear as orange points. While the narrow bin data appears a factor of 1.5$-$2 higher than that of the wide bins in the w($\theta$) plots (top), we show that converting this into w$_p$(r$_p$) units shows little discrepancy between the separate narrow redshift ranges and the wider redshift-- (the dashed red line marks where the narrow and wide bins give equivalent results. In most cases, the  correlation function derived using wide z bins corresponds to the average correlation function of the narrower bins. Therefore, selecting a dz$=$0.4 redshift bin will not systematically affect the correlation strength.}
   \label{fig:acor_all}
\end{figure}

 \begin{figure}[h]
   \centering
   \includegraphics[width=6.5in]{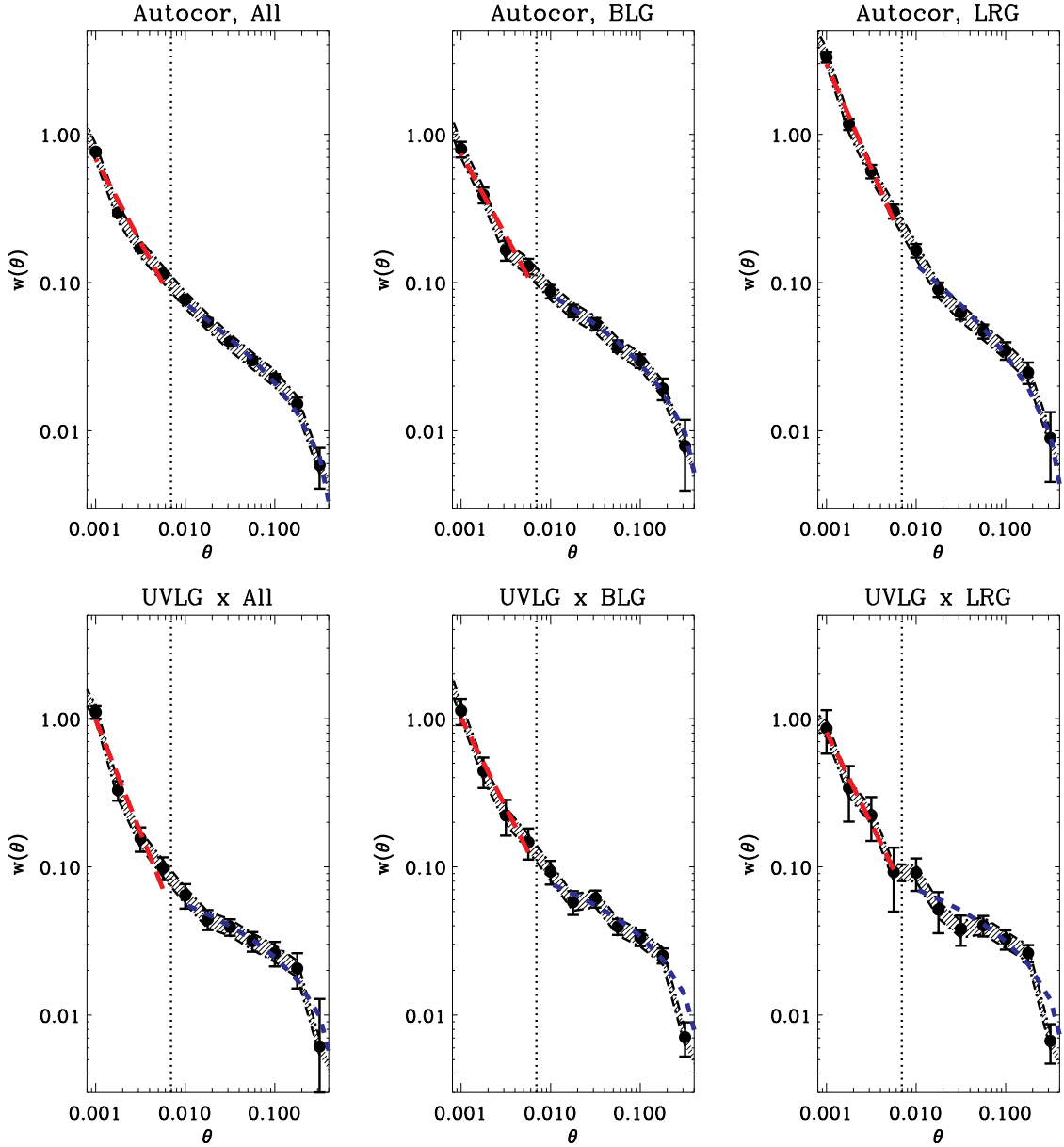} % requires the graphicx package
   \caption{Fits to the angular correlation function, for both auto-corrlelations (top) and cross-correlations with UVLGs (bottom) for the ALL (left), BLG (middle) and LRG (right) samples. The shaded regions within the dashed lines provide errors caused by cosmic variance. The dotted vertical line delineates the physical scale of 200 kpc, the critical distance to separate large scale from small scale. The blue dashed lines show the large scale fit, including the IC correction to the data, while the red long-dashed lines mark the best fit to the small scales. Again, we demonstrate our fits only for one representative redshift sample: 0.8$-$1.2. }
   \label{fig:fit}
\end{figure}

 \begin{figure}[h]
   \centering
   \includegraphics[width=4.5in]{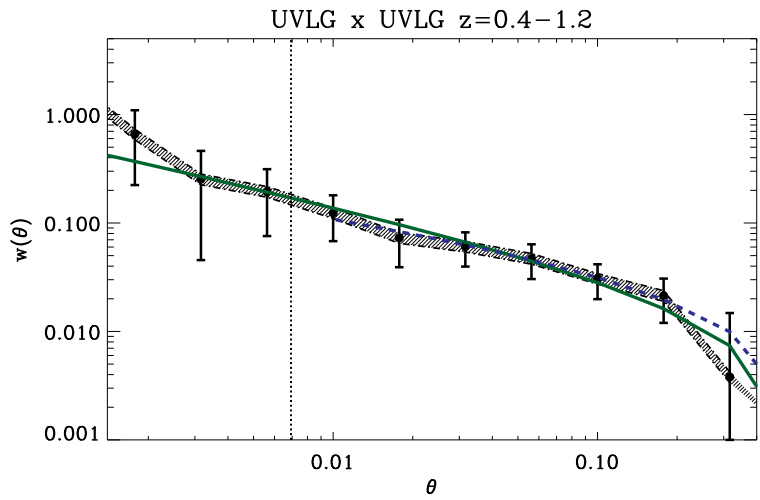} % requires the graphicx package
   \caption{Fits to the angular correlation function, for auto-corrlelations of UVLGs. The shaded regions within the dashed lines provide errors caused by cosmic variance. The dotted vertical line delineates the physical scale of 200 kpc, the critical distance to separate large scale from small scale. The blue dashed line shows the large scale fit, including the IC correction to the data, and green solid line marks the best fit to the entire range of scales. }
   \label{fig:fituvlg}
\end{figure}

\begin{figure}[h]
   \centering
   \vspace{-0.2in}
   \includegraphics[height=6.5in]{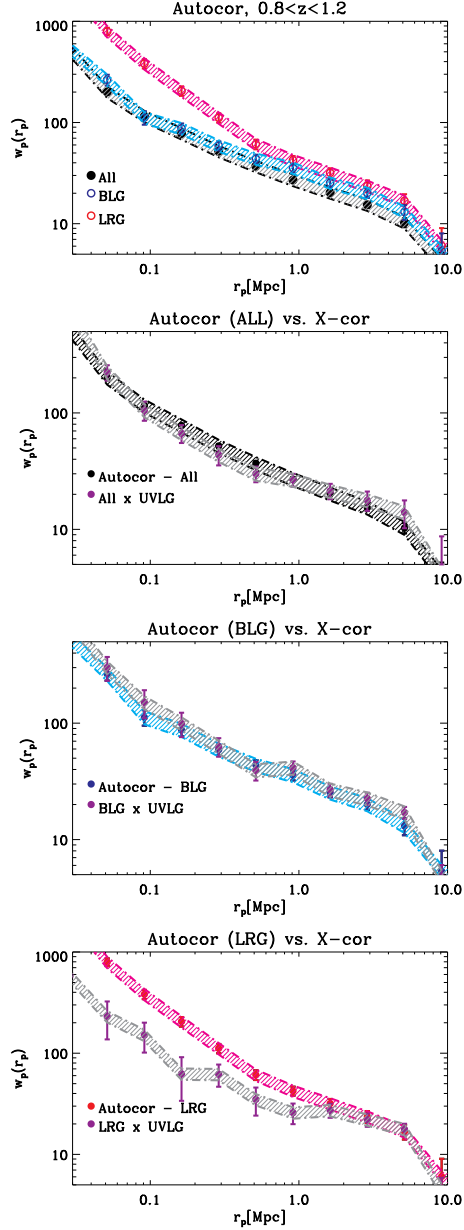} 
   	\vspace{-0.2in}
   \caption{Comparing correlation functions for our samples, for the 0.8$-$1.2 redshift data. The shaded regions within the dashed lines provide errors caused by cosmic variance.  The topmost panel compares auto-correlation of the ALL sample (black) with BLG sample (open navy circle) and LRG sample (open red circle). The second panel displays the ALL auto-correlation (black) with the `ALL x UVLGs' cross-correlation (purple). The third panel, similarly, shows the BLG auto-correlation (blue)  in comparison with the `BLG x UVLGs' cross-correlation (purple). Finally, in the bottom panel the auto-correlation of `LRGs' is shown in red, with the cross-correlation of `LRGs' with UVLGs  shown in purple. While on small scales, the auto-correlations of the ALL and BLG samples are indistinguishable from their cross correlations with the UVLGs, the LRGs appear to cluster with themselves more strongly than with UVLGs. }
   \label{fig:acor_comp}
\end{figure}

 \begin{figure}[h]
   \centering
   \includegraphics[height=4in]{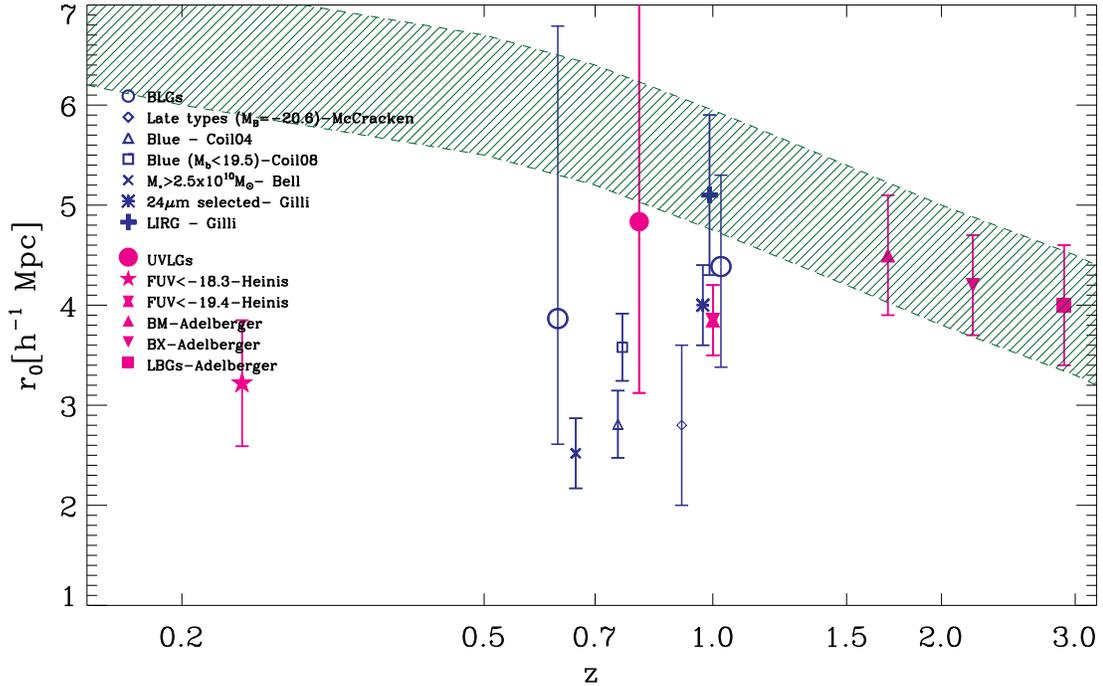} % requires the graphicx package
   \caption{UVLG correlation length, determined from the autocorrelation of 0.4$<$z$<$1.2 UVLGs. We compare with correlation length data for star-forming galaxies from other studies. We also include high redshift results for LBGs, BX, and BM (see \cite{adel05} for their description) with low-redshift \GALEX results to compare UVLGs with other UV-selected samples (all shown in magenta). The dark green shaded region shows the evolution of a 10$^{11.2}-$10$^{11.8}$ M$_\odot$ halo with redshift (identical to \cite{adel05}).}
   \label{fig:r0_uvlg}
\end{figure}

\begin{figure}[h]
   \centering
   \includegraphics[height=7.5in]{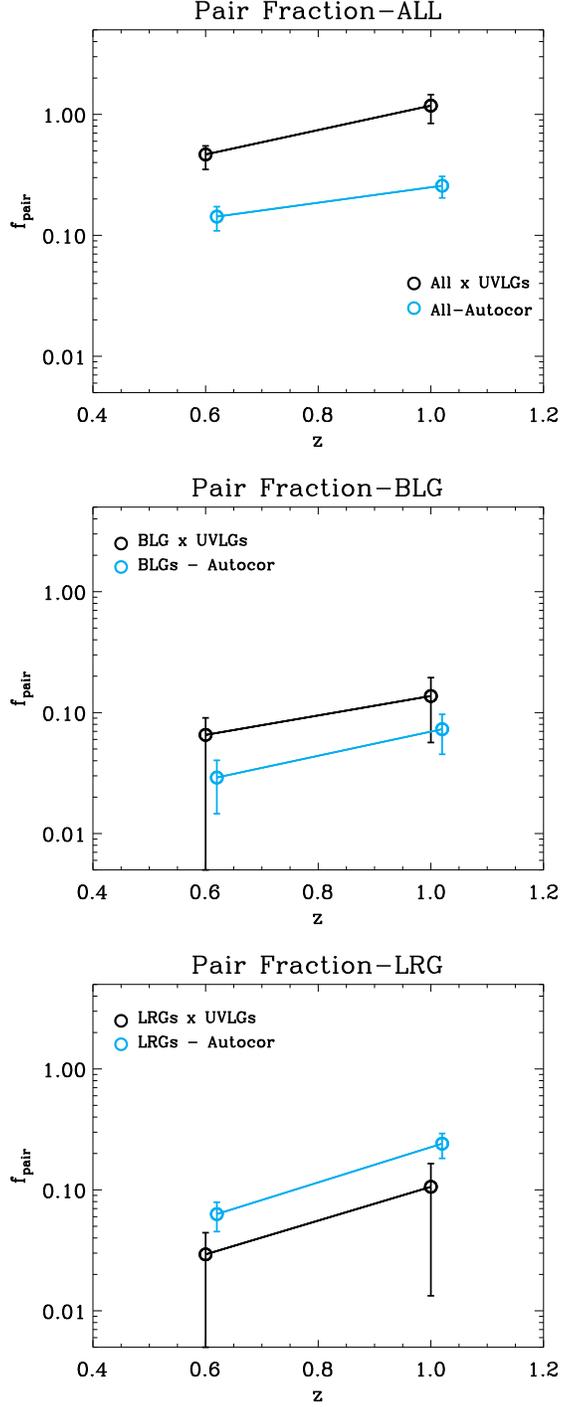} % requires the graphicx package
   \caption{Evolution of pair fractions--Top panel: pair fraction of ALL sample with itself (cyan) compared to ALL with UVLGs (black). UVLGs have an increased likelihood of interacting with ALL galaxies than the ALL sample interacting with itself. Middle panel: BLGs interact with themselves (cyan) similarly to UVLGs interacting with BLGs (black). Finally,  LRGs are marginally more likely found in pairs with other LRGs (cyan) compared to being found near UVLGs. }
   \label{fig:comp}
\end{figure}
%
%\begin{figure}[h]
%   \centering
%   \includegraphics[width=6in]{fig10.eps} % requires the graphicx package
%   \caption{Comparison of small-scale clustering for ALL auto-correlation (black points), UVLG-ALL cross-correlation (purple points) or LBA-ALL cross correlation (orange points), shown with their fits in this small scale regime. The shaded regions within the dashed lines provide errors caused by cosmic variance. LBAs appear to pair with the ALL sample, similar to UVLG-ALL and ALL-ALL.}
%  \label{fig:lba}
%\end{figure} 

\begin{figure}[h]
   \centering
   \includegraphics[height=4.5in]{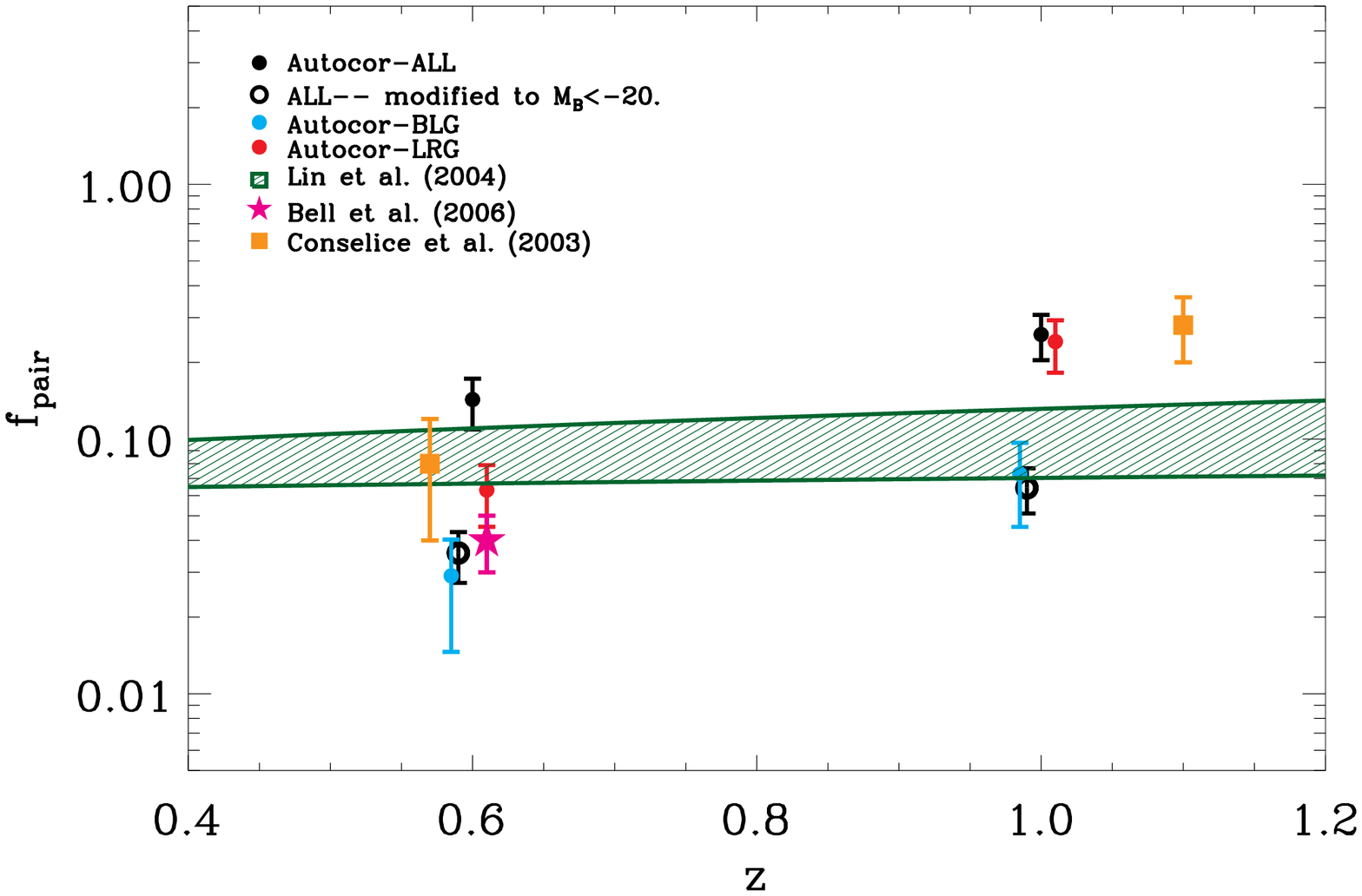} % requires the graphicx package
   \caption{Comparing auto-correlation results for the three comparison samples-- ALL(black points), BLGs (blue points), and LRGs (red points) with other published results. \cite{Lin04} (green shaded region) studies the evolution of pair fraction using galaxies with $-22\le$M$_{\rm B} \le-20$; \cite{Bell06} (magenta star) studies a sample with M$_{\rm B}<-20$; \cite{Conselice903} (orange square) finds merger fraction, assumed to be half the pair fraction, for M$_{\rm B}<-18$. These magnitude limited samples, with no color selection, are most comparable to our ALL sample. From Eqn. \ref{eqn:pf} we see that the pair fraction depends on number density, which depends strongly on limiting magnitude. In order to compare our pair fraction with other studies, we extrapolate the pair fraction of the ALL sample for M$_{\rm B}< -20$, changing only the number density, and assuming that the correlation lengths is the same as our ALL sample (See discussion in \cite{Bell06}). These points are shown as open circles, and appear to agree well with the other studies.}
   \label{fig:pfpub}
\end{figure}

\end{document}